\documentclass[sigconf,natbib=false]{acmart}
\usepackage{multirow}
\usepackage{booktabs}
\usepackage{graphicx}
\usepackage{url}
\usepackage{hyperref}

\AtBeginDocument{%
  }

\setcopyright{acmlicensed}
\copyrightyear{2026}
\acmYear{2026}
\acmDOI{XXXXXXX.XXXXXXX}

\acmConference[EASE 2026]{International Conference on Evaluation and Assessment in Software Engineering}{June 9--12, 2026}{Glasgow, United Kingdom}
\acmISBN{978-1-4503-XXXX-X/2018/06}

\RequirePackage[
  datamodel=acmdatamodel,
  style=acmnumeric,
 ]{biblatex}

\begin{document}

\title [Does social identity matter in software engineering?]{Does social identity matter in software engineering?  Assessing~the case of research software engineers}

\author{Chukwudi Uwasomba}
\orcid{0000-0002-6572-2835}
\affiliation{%
\institution{The Open University}
\city{Milton Keynes}
\state{England}
\country{UK}
}
\email{chukwudi.uwasomba@open.ac.uk}

\author{Tamara Lopez}
\affiliation{%
  \institution{The Open University}
  \city{Milton Keynes}
  \state{England}
  \country{UK}}
\email{tamara.lopez@open.ac.uk}

\author{Melanie Langer}
\affiliation{%
  \institution{INSEAD}
  \city{Fontainbleau}
  \country{France}
}
\email{melanie.s.langer@gmail.com}

\author{Helen Sharp}
\affiliation{%
  \institution{The Open University}
  \city{Milton Keynes}
  \state{England}
  \country{UK}}
\email{helen.sharp@open.ac.uk}

\author{Michel Wermelinger}
\affiliation{%
  \institution{The Open University}
  \city{Milton Keynes}
  \state{England}
  \country{UK}}
\email{michel.wermelinger@open.ac.uk}

\author{Caroline Jay}
\affiliation{%
 \institution{The University of Manchester}
 \city{Manchester}
 \state{England}
 \country{UK}}
\email{caroline.jay@manchester.ac.uk}

\author{Mark Levine}
\affiliation{%
  \institution{Lancaster University}
  \city{Lancaster}
  \state{England}
  \country{UK}}
\email{r.levine@lancaster.ac.uk}

\author{Bashar Nuseibeh}
\affiliation{%
  \institution{The Open University}
  \city{Milton Keynes}
  \state{England}
  \country{UK}}
\email{bashar.nuseibeh@open.ac.uk}


\renewcommand{\shortauthors}{Uwasomba et al.}

\begin{abstract}
Social identity is a concept from psychology that refers to the part of an individual’s identity that derives from their group membership(s). In this paper, we explore social identity in members of the professional community of Research Software Engineers (RSEs). Using a mixed-methods approach, our study combined computational linguistic analysis and inferential statistics to examine over 28,000 social media posts, 1,700 blogs, and survey responses from 381 professional RSEs. The findings highlight the emergence of a collective RSE identity and demonstrate its role in shaping professional wellbeing. This study contributes an interdisciplinary perspective by integrating social psychology and software engineering to show how a professional identity evolves and why it matters.\\
\end{abstract}

\ccsdesc[500]{Software and its engineering~Software organization and properties~Software development techniques}
\ccsdesc[300]{Human-centered computing~Collaborative and social computing~Empirical studies in collaborative and social computing}
\ccsdesc[300]{Social and professional topics~Professional topics~Computing occupations}
\ccsdesc[200]{Applied computing~Psychology}
\ccsdesc[200]{Applied computing~Sociology}

\keywords{Social identity, Research Software Engineers, Resilience, Computational
linguistics, Inferential statistics}

\maketitle

\section{Introduction}

Professional developers work within a distinct community and culture of practice \cite{sharp2002software}.  The social aspects of software development have been studied in a number of ways: in terms of developer experience \cite{fagerholm_developer_2012, greiler_actionable_2022}; and attributes of work that keep developers motivated \cite{sharp_models_2009}, satisfied \cite{França2020} and productive \cite{Storey2022}. Among other findings, these kinds of studies identify features that make a workplace attractive to software engineers \cite{moe_attractive_2023}, that will prevent turnover \cite{kuutila2025staying}, and that may help a software engineer develop in their career \cite{hochstein_developing_2023}.  

Despite a focus on social aspects of developer experience, social identity has received limited attention. This is a significant research gap. Social identity refers to the part of an individual’s identity that derives from their group membership(s). These social identities shape individuals’ everyday interactions and have far-reaching implications for nearly every aspect of human experience, including attitudes \cite{Reynolds2001}, cognition \cite{Haslam1999}, emotions \cite{Doosje1998}, behaviour \cite{Shih1999}, and actions and interactions \cite{Hogg2006}.

Baeckevik et al. \cite{Baeckevik2019} examined software engineers' social identity to understand how group identities relate to team effectiveness. Although this small study found an interplay between their personal identity and private and public self-esteem, it did not manage to fully capture connections between these factors and team effectiveness. There are also clusters of studies within software engineering (SE) that examine concepts similar to social identity. For example recent work looked at perceptions of success and belonging within open source communities \cite{trinkenreich_pots_2021} and companies \cite{trinkenreich_unraveling_2024}. 

Addressing a call for larger scale examination of social identity within SE \cite {Baeckevik2019}, this research focuses on one group of professional software engineers. Building upon research and professional experience within the authoring team, and capitalising on its recent formation \cite{RSESociety2021}, we chose research software engineers (RSEs). RSEs are professional software engineers who support different research domains, e.g. physics, medicine, climate change or biology \cite{segal2008developing}. Some elements of this group have been studied by the research community \cite{carver2007software} and discussed at workshops \cite{ICSE2008workshop}, but not in terms of their social identity.  

This two-part study examines the RSE profession to establish whether social identities in software engineering communities matter and are worthy of study, asking:\\
\\
\textbf{RQ1} How do social identities cohere over time within software engineering?\\
\\
\textbf{RQ2} How does orientation toward a social identity influence professional outcomes within software engineering?\\
\\
To answer these questions, we established an empirical basis and method for tracing the collective evolution of the RSE social identity through discursive patterns in public data collected from over 28,000 social media posts and 1,700 blog posts. In addition, we collected survey data from 381 software engineers working within research environments. The data were analysed using psychological constructs that focus on how someone responds to "threats" in the workplace that have been linked with professional outcomes, resilience and autonomy \cite{Manning2003} \cite{Bledow2021}.

Using the case of research software engineering, this work provides evidence that stronger identification with a social identity enhances professional outcomes and gives insight into the role that autonomy and resilience play as mediators between social identity and positive professional outcomes.

\section{Literature Review}
This section gives background (\ref{sec:research-software-engineers}) to the research software engineering field; (\ref{sec:social-identity-collective-action}) outlines collective action, a theory in social identity; (\ref{sec:threat-attachment-style-action-state}) briefly explains attachment theory and action-state orientation; and (\ref{sec:outcomes}) defines professional outcomes as they are studied in software engineering and other fields.

\subsection{Research Software Engineers}
\label{sec:research-software-engineers}

The term Research Software Engineer was first coined at a Collaborations Workshop at Queen’s College Oxford on March 21st, 2012, as part of a movement working towards the recognition and advancement of the role and of those engaged in software engineering to support research \cite{Hettrick2016} \cite{RSESociety2021}. In 2014, while 400 out of 10,000 academic job advertisements were related to software development, these 400 used almost 200 different job titles \cite{Hettrick2016}. The lack of a clear professional identity went hand in hand with a lack of career path, and a lack of opportunity for progression and job security.

To move beyond being seen as  as auxiliary to research, it was determined that ``Research Software Engineers needed a community,'' and that “to make the[ir new governing] association sustainable,'' that association ``needed to be community led" \cite{Hettrick2016}. The identity itself was critical to this development. As Ian Cosden, the director of Princeton University’s Research Software Engineering group, said ``it’s not really the role that’s completely new… it’s the formality, it’s the awareness, it’s the title'' \cite{OCarroll2021}.
Prior to the creation of a formal community, ``many (RSEs) thought they were the only person conducting this highly valued but unrecognized work. The strength of the UKRSE Association (the first formal RSE group) is that it shows RSEs they are not alone'' \cite{Hettrick2016}. 

Articulating and organizing around a shared social identity has allowed these professionals to create research software groups within universities that are substantial enough to provide a promotion path that sustains long-term careers. The development of the RSE identity and associated structures have benefited other academics in the RSEs’ organizations as well, providing ``a flexible and cost-effective service for researchers” \cite{Hettrick2016}. 
As Cosden points out, ``researchers who can collaborate with RSEs are going to be able to do things that others are not” \cite{OCarroll2021}. Their contribution is critical to disciplinary advances, for example to model the spread of the covid-19 virus  \cite{marion2022modelling}, but recognition of their role has been slow in coming.  Despite the indispensable nature of their contributions, RSEs have faced a lack of clear career pathways, opportunities for progression, job security and status \cite{Parr2013} \cite{Smith2014} \cite{PawlickaDeger2022}.

\subsection{Social Identity and Collective Action}
\label{sec:social-identity-collective-action}
Social identity is a key psychological construct that refers to the part of an individual’s identity that derives from their group membership(s) \cite{Tajfel1979} \cite{Turner1987}. 
The effects of social identification are present in many fields, including economics, education, political science and sociolinguistics \cite{Reicher2010}. Social identities can include professional and organizational identities, and the interplay of the various identities held by individual employees impacts performance, cooperation, conflict, and wellbeing \cite{Haslam2004} \cite{Steffens2017} \cite{Wegge2014}. 
Social identities have a positive effect on mental health and wellbeing, even being referred to as a “social cure” \cite{Jetten2012} \cite{Haslam2018}. Social identification with more groups increases these positive effects \cite{Haslam2016}. Of particular relevance for this paper, these effects include increased resilience \cite{Haslam2008} \cite{SeymourSmith2017}. 

Group identification is fundamental for the value that the group contributes to society to be recognized, and for that value to be fairly compensated, including through inclusion and respect. Rational choice perspectives \cite{Downs1957} \cite{Riker1968} 
have  been used to understand an individual’s decision to participate in collective action, reflecting the economic nature of the decision in weighing the potential gains against the potential costs, and their respective probabilities with and without one’s own participation \cite{Muller1986}. However, more recent research has stressed the importance of social identification, anger, and a sense of efficacy.  Through social identification, an individual comes to understand the circumstances that are negatively affecting them as shared with others in their social group \cite{Jost2012} \cite{Kelly1996}  \cite{Subasic2008}. The combination of this process with the perception of injustice and a resulting sense of anger \cite{VanZomeren2004} \cite{Wakslak2007}, as well as an assessment of the group as capable of achieving its goal, i.e. having collective efficacy \cite{VanZomeren2012}, form the basis for the members of a group to take part in collective action to improve their shared circumstances.

\subsection{Threat, Attachment Style, and Action-state Orientation} 
\label{sec:threat-attachment-style-action-state}
In the workplace, the need to manage various forms of threat can negatively and positively motivate behavior, leading to a fear of making individual mistakes \cite{Boyes2020}, or losing one's job \cite{Prentice2022}. Other threats within the workplace include loss of perceived status, respect, and value to others in an organization \cite{Blader2017}. 

Identification with one’s social group(s) 
offers substantial self-protective advantages against threats \cite{Jetten2012}\cite{Haslam2018}. While there are numerous individual differences related to how someone may respond to threat, many capture some aspect of a tendency to orient towards versus away from threat. This study focuses on attachment style \cite{Bowlby1969} \cite{Ainsworth1978} \cite{Mikulincer2003a} and action-state orientation \cite{Kuhl2000} \cite{Diefendorff2000}. 

Attachment styles reflect individual differences in orientations toward close others and toward potential threats in the environment, which may motivate or inhibit interpersonal behavior \cite{Mikulincer2003b}. An \textit{anxious} attachment style refers to the interpersonal tendency to focus on one’s relationships and concerns about them, and to exaggerate problems in a way that motivates attempts to engage close others in providing protection and comfort. An \textit{avoidant} attachment style, conversely, refers to the tendency to direct attention away from one’s relationships and relationship-related distress, as well as from potential threats that might motivate attempts to seek protection and comfort from relationship partners. These two attachment styles are referred to as \textit{insecure}, whereas a \textit{secure} style refers to comfortably accepting intimacy and support, without exaggerating motivation or behavior to obtain them. 
Attachment style has been linked with more and less positive professional outcomes \cite{Manning2003}, as well as theoretically related individual differences in cognitive style \cite{Mikulincer1997}. In particular, attachment style has been associated with levels of job satisfaction \cite{Toepfer1996}, 
resilience, stress, and coping \cite{Klohnen1998} \cite{Caldwell2012}. 

Individual differences in action-state orientation describe a similar distinction in strategies for managing threat, but without the interpersonal focus. As for anxious attachment, a state orientation entails orientation towards threat and a focus on the problem itself, whereas an action orientation entails orientation away from threat, and instead a focus on what actions an individual can take themselves \cite{Bledow2021}.  In a management context \cite{Diefendorff2000} and specifically for creative problem-solving, different action-state orientations result in better outcomes depending on the extent of one’s professional autonomy within an organisation \cite{Bledow2021}.

\subsection{Professional Outcomes}\label{sec:outcomes}
Professional outcomes refer to a set of achievements that an individual attains through their work. The term can encompass a wide range of issues but this research focuses on four outcomes that are known to be important to software engineers: success, job satisfaction, meaningfulness and team communication. 

\textit{Success} is a subjective measure and feeling successful in work life is recognised as motivational \cite{Ryan2017}. However, \textit{success} may mean different things to different people, especially if they are in different project roles. For example software engineers regard individual success in terms of doing a good job, having a sense of achievement, professional growth and learning \cite{procaccino2005}. For RSEs, recognition of their contributions to research software is important \cite{RSEcredit}. 

\textit{Job satisfaction} has been a mainstay of management and organizational research \cite{Brooke1988} \cite{Cano2004}. From the early days of computer-based work, job satisfaction and motivation of IT workers have been the subject of many studies \cite{JCT1991}\cite{Sharma2017}. 
It remains an active focus in software engineering, and among other findings, it has been shown that satisfied engineers are happy \cite{Franca2014}, and that happiness can impact their productivity and wellbeing \cite{Graziotin2019}. 

\textit{Meaningfulness} is a key area of organizational research \cite{Rosso2010}, defined as ''a subjectively meaningful experience consisting of experiencing positive meaning in work, sensing that work is a key avenue for making meaning, and perceiving one’s work to benefit some greater good” \cite{Steger2012}. Meaningfulness has been found to be associated with greater well-being \cite{Arnold2007}, and with feeling one’s work has a higher level of value and importance \cite{Nord1990} \cite{Harpaz2002}. This is particularly relevant if the development of the RSE social identity is viewed as an instance of collective action meant to benefit RSE workers. 

Good \textit{team communication}, defined as ''sharing of information, ideas, and knowledge between two or more team members” \cite{uwasomba2016managing}, is fundamental to teamwork. Prior work has demonstrated that successful team communication is important to achieving positive outcomes in emergency or crisis situations \cite{Ghaferi2016} \cite{Davis2017}, which has a relationship with resilience \cite{Hollnagel2014}. RSEs typically collaborate in project teams, both with other RSEs and with non-RSE academics within their scientific disciplines.

\section{Hypotheses}
\label{sec:hypotheses}
The recently formed research software engineering group offers an opportunity to examine the development of a social identity in a professional domain.  Based on the literature reviewed in the previous section, we specify the following hypotheses. The full set is summarised in Table \ref{tab:aims_hypotheses_analysis2}.

Social media and online forums have become a valuable source of language data for social scientists studying the development of social identities and their role in collective action  \cite{Langer2019, back2018we}. This literature has demonstrated the utility of online text data for both detecting identities \cite{koschate2021asia} and tracking their formation, including specifically in an ideological or activist context \cite{back2018we}.  Prior research has shown that growing ingroup identification can be charted through the use of pronouns that signal individuals’ perception of a meaningful and coherent group and of their membership in it \cite{back2018we}, or group differentiation \cite{gustafsson2014selection}.

We expected that, as demonstrated by \cite{Smith_social_2015, SmithThomasMcGarty2015,  VanZomeren2012, Langer2019}, within public discourse about RSE there will be increasing evidence in social media and online forums of a shared social identity (\textbf{H1}), decreasing evidence over time of the emotions and cognitive processes associated with collective action (\textbf{H2}), increasing evidence of a shared linguistic style associated with the group (\textbf{H3}), and a prevalence of topics associated with the group and its formation (\textbf{H4}). In relation to \textbf{H1} we also expected that as a shared social identity develops, the use of:
  \begin{description}
    \item[H1a:]  singular first-person pronouns (eg. \textit{I, me}) will decrease. 
    \item[H1b:]  plural first-person pronouns (eg. \textit{we, us}) will increase.  
    \item[H1c:]  plural third-person pronouns (eg. \textit{they, them}) will increase.
  \end{description}

As the RSE identity develops and moves farther from the start of the collective construction of the identity \textbf{H2}, social media posts will contain:
  \begin{description}
    \item[H2a:] less negative emotion (eg. \textit{bad, hate, hurt}). 
    \item[H2b:] less anger (eg. \textit{angry, mad, frustrated}).
    \item[H2c:] less focus on the past (eg. \textit{was, had, been}). 
    \item[H2d:] less discrepancy (eg. \textit{would, could, want}).
    \item[H2e:] less differentiation (eg. \textit{but, not, if}). 
  \end{description}

 The need to manage various forms of threat is an important motivator of human behavior in individuals’ professional lives. It can also cover threats to perceived status, respect, and value to others in the organization \cite{Blader2017}. Identification with one’s social group(s) has been found to offer substantial self-protective advantages \cite{Haslam2018}. Given social identification’s positive effects on mental health and the centrality of collective efficacy to participation in the RSE movement \cite{OCarroll2021}, we anticipated higher levels of social identification to be associated with improved professional outcomes, as well as with both autonomy and resilience \cite{Manning2003} \cite{Bledow2021}. 
 
 Specifically, we expected that stronger social identification with the RSE identity will predict: \textbf{H5} improved self-reported professional outcomes (Section \ref{sec:outcomes});  \textbf{H6} greater self-reported feelings of autonomy; and \textbf{H7} greater self-reported feelings of resilience. 

We furthermore expected autonomy and resilience to each mediate the relationship between social identification and professional outcomes: \textbf{H8} The relationship between social identification with the RSE identity and self-reported satisfaction with professional outcomes will be mediated by self-reported feelings of autonomy (\textbf{H8a}) and of resilience (\textbf{H8b}). 

Following the research linking attachment security with positive outcomes (Section  \ref{sec:threat-attachment-style-action-state}), we expected
:  \textbf{H9} Attachment security will have a positive relationship with social identification(\textbf{H9a}), autonomy (\textbf{H9b}), and resilience (\textbf{H9c}). 


\begin{table*}[h!t]
\footnotesize
\caption{Mapping of Research Aims, Research Questions, Hypotheses, Support, and Analyses}
\centering
\label{tab:aims_hypotheses_analysis2}
\renewcommand{\arraystretch}{1.2}
\begin{tabular}{|p{3.5cm}|p{3cm}|p{3cm}|p{3.5cm}|p{2.5cm}|}
\hline
\textbf{Research Aims} & \textbf{Research Questions} & \textbf{Hypotheses} & \textbf{Hypothesis Support} (yes $\bullet$, no $\circ$) & \textbf{Analysis Conducted} \\ \hline
\multirow{9}{3.5cm}{%
    Through an analysis of Research Software Engineering (RSE), learn more about how group social identities form, and examine how identification with a particular social group relates to professional outcomes.

} & 
    \multirow{4}{3cm}{%
         RQ1. How do social identities cohere over time within software engineering?%
    } &
    H1(a-c): Pronoun use & Twitter: $\bullet\bullet\bullet$ ~Blog: $\bullet\bullet\circ$
     &
    \multirow{4}{2.5cm}{%
        Linguistic style analysis. Topic modelling. Principal components analysis (PCA)%
    } \\ \cline{3-4}
& & H2(a–e): Emotional tone & Twitter: $\bullet \bullet \circ \bullet \circ$
~Blog: $\bullet\bullet\bullet\circ\bullet$ & \\ \cline{3-4}
& & H3: Linguistic style & Twitter/Blog: $\circ$ & \\ \cline{3-4}
& & H4: Discussion topics & Twitter/Blog: $\bullet$ & \\ \cline{2-5} 
& \multirow{4}{3cm}{%
        RQ2. How does orientation toward a social identity influence professional outcomes within software engineering?%
    } &
    H5: Professional outcomes & 
    $\bullet$ &
    \multirow{4}{2.5cm}{%
        Descriptive statistics. Inferential statistics. Mediation analyses%
    } \\ \cline{3-4}
& &  H6: Autonomy & $\bullet$ & \\ \cline{3-4}
& &  H7: Resilience & $\bullet$ & \\ \cline{3-4}
& & H8(a-b): Mediation effects & 
$\bullet ~\bullet$  & \\ 
\cline{3-4}
& & H9(a-c): Social identification & $\circ \circ \circ$ & \\ \hline
\end{tabular}

\end{table*}

\section{Method}
To assess the social identity of RSEs, this study adopted a mixed-methods design \cite{robson2016real}, combining computational analyses of social media and blog posts, with inferential statistical analysis of survey data.  The methods provide a counterpart to one another: computational techniques provide a way to establish the boundaries around social identity processes, while self-reported data within surveys provide access to complex and nuanced attitudes of individuals. The study was preregistered on the Open Science Framework (details available in \hyperref[sec:supplementary-materials]{Supplementary Materials}).

The following subsections contain two parts: the first relates to the application of computational analyses to social media and blog data, and the second, to the collection and analysis of survey data.

\subsection{Research Sample and Datasets}

\subsubsection{RSE-affiliated online discourse} To support computational linguistics analyses, 
28,412 tweets were identified in 19 RSE-affiliated Twitter accounts (Table \ref{tab:twitter_accounts}) and 1,765 blog posts made by six major RSE-related websites (Table  \ref{tab:blogs_overview}). These digital texts represent the collective discourse of the RSE community from January 2012 to the date of data collection in 2022, encompassing the inception and maturation of the professional identity. The websites and Twitter accounts belong to organizations that identify as RSE organizations or are focused on RSEs. The accounts were identified through searching Twitter for terms related to research software engineering (e.g., rse, research software engineer, research software, researchsoftware, etc.), examining users who had interacted with posts by other organizations or related to known RSE events, and searching for known RSE organizations. Websites were found by searching the internet for these terms on the recommendations of RSE colleagues and leaders. The searches for Twitter accounts and websites were mutually reinforcing, i.e. we explored whether Twitter accounts had associated websites and vice versa. For example, The Software Sustainability Institute describes itself on its website as “Cultivating research software to support world-class research,” and on its Twitter account, @SoftwareSaved, by saying “We provide resources and expertise to cultivate world-class research with software”.

\begin{table}[h!t]
\footnotesize
\caption{Overview of Twitter Accounts Used in the Study}
\centering
\label{tab:twitter_accounts}
\renewcommand{\arraystretch}{1.2}
\begin{tabular}{|c|p{2.2cm}|r|r|r|}
\hline
\textbf{\#} & \textbf{Twitter Account} & \textbf{Scraped} & \textbf{Cleaned} & \textbf{Removed Retweets} \\
\hline
1  & TW01        & 7323  & 7320  & 3   \\ 
2  & TW02      & 2358  & 2353  & 5   \\ 
3  & TW03             & 111   & 111   & 0   \\ 
4  & TW04            & 18    & 18    & 0   \\ 
5  & TW05             & 701   & 693   & 8   \\ 
6  & TW06           & 292   & 292   & 0   \\ 
7  & TW07          & 162   & 162   & 0   \\ 
8  & TW08            & 493   & 493   & 0   \\ 
9  & TW09              & 336   & 336   & 0   \\ 
10 & TW10            & 363   & 363   & 0   \\ 
11 & TW11         & 953   & 952   & 1   \\ 
12 & TW12          & 100   & 100   & 0   \\ 
13 & TW13       & 7460  & 7460  & 0   \\ 
14 & TW14       & 661   & 661   & 0   \\ 
15 & TW15             & 980   & 980   & 0   \\ 
16 & TW16            & 546   & 425   & 121 \\ 
17 & TW17                & 5126  & 5124  & 2   \\ 
18 & TW18        & 927   & 426   & 501 \\ 
19 & TW19         & 144   & 143   & 1   \\ 
\hline
\multicolumn{2}{|l|}{\textbf{Total Tweets}} & \textbf{29054} & \textbf{28412} & \textbf{642} \\
\hline
\end{tabular}
\end{table}

\begin{table}[h!t]
\footnotesize
\caption{Overview of Blogs Used in the Study}
\centering
\label{tab:blogs_overview}
\renewcommand{\arraystretch}{1.2}
\begin{tabular}{|p{5.5cm}|r|r|}
\hline
\textbf{Blog Name} & \textbf{Scraped} & \textbf{Cleaned} \\
\hline
Research Software Engineers International & 10 & 10 \\
Better Scientific Software (BSSw)         & 102 & 102 \\
Software Sustainability Institute         & 1137 & 1137 \\
RSE Sheffield Blog                        & 96 & 96 \\
Research Software Alliance Blog           & 4 & 4 \\
University of Manchester IT Blog          & 416 & 416 \\
\hline
\textbf{Total blog posts}                 & \textbf{1765} & \textbf{1765} \\
\hline
\end{tabular}
\end{table}

\subsubsection{Professional RSEs} 381 survey participants (Mean age = 37.58, SD = 9.21) were recruited from professional networks of RSEs. The sample comprised 134 women, 190 men, 5 non-binary participants, and 2 participants who preferred not to disclose their gender. Participants responded to a recruitment notice disseminated by the authors and by RSEs in the authors’ professional network via Twitter, LinkedIn, the Software Sustainability Institute’s (SSI) SSI Fellows listserv, a blogpost on the SSI website, a post on the SSI’s Twitter account, the UK RSE Slack Group, research software engineers at King's College London, Indiana University, and McGill University, the code4lib listserv, and the Samvera technical group. In addition, the recruitment notice asked individuals to share the notices with anyone in their “circles who does software engineering work in research.” Eligibility required active involvement in research-related software development. Participants were given an Amazon voucher (£4.50) to make participation accessible across geographic regions.
\subsection{Data Collection}
\label{section3.3}
\subsubsection{Social Media and Blog Posts} Data from Twitter were scraped using Twitter’s API v2 in a custom Python script available on GitHub, which retrieved all accessible tweets, and saved them in a CSV file. Blog posts were scraped using Google Chrome and its browser extension webscraper.io and saved in CSV format. This required the creation of a scraper configuration file for each blog, which identified parts of the blog to extract: the title, text corpus, author(s).  Twitter and blog datasets were processed separately to account for differences in text length and formatting. 

The scraped data of each Twitter account were normalized and cleaned using custom R-scripts. They used the \texttt{textclean} library and custom functions to remove or replace text that might interfere with LIWC-22, the text analysis software ~\cite{Boyd2022}. For example, emojis and emoticons where replaced with word form equivalents; abbreviations were replaced with their standard word form; HTML tags, line breaks, extra spaces, photo credits were removed. Blog posts were also pre-processed using R-scripts tailored to their structure.  Following pre-processing, the Twitter and blog datasets were merged into separate CSV files for analysis, one containing tweets and one containing posts. 

\subsubsection{Survey Instrument}
\label{subsubsec:Survey-data-collection}Data were collected from participants in an online survey platform. Informed consent was obtained prior to participants' involvement. Participants were asked to provide demographic and background information such as age, number of years writing software, country of employment, job title, and professional environment (e.g. industry or university). They were then asked to indicate their agreement with the statements in Table \ref{tab:instrument} on a scale of 1 (Strongly Disagree) to 7 (Strongly Agree). Statements reflect different aspects of each construct of interest: the professional outcomes defined in Section \ref{sec:outcomes}, i.e. success, satisfaction, meaningfulness and team communication, Breaugh's three categories of job autonomy \cite{Breaugh1985}, Hollnagel's four cornerstones of resilience \cite{Hollnagel2011}, and attachment style. 
We chose academic authorship \cite{allenetal2017} to assess success from the RSE perspective; borrowing from \cite{Sharma2017} job satisfaction was investigated in terms of pay, training and promotion; meaningfulness is subjective and so was assessed through direct statements; team communication focused on the frequency and openness of inter-team interactions. 

\begin{table}[ht]
\footnotesize
\caption{Survey Instrument}
\label{tab:instrument}
\centering
\renewcommand{\arraystretch}{1.2}
\begin{tabular}{|p{2.5cm}|p{5.0cm}|}
\hline
\textbf{Construct} & \textbf{Likert scale statements related to the construct} \\
\hline

\multirow{4}{5.8cm}{RSE Work} 
  & My work involves writing or developing software. \\
\cline{2-2}
  & I am considered the/a ‘go-to’ person when people have questions about writing code or developing software. \\
\cline{2-2}
  & People ask my advice about writing code/ developing software. \\
\cline{2-2}
  & Software engineering is an explicit part of my job description. \\
\hline

\multirow{6}{5.8cm}{Social Identity} 
  & I have a lot in common with the average RSE. \\
\cline{2-2}
  & Members of this group have a lot in common. \\
\cline{2-2}
  & It gives me a good feeling to be an RSE. \\
\cline{2-2}
  & I feel solidarity with RSEs. \\
\cline{2-2}
  & Being an RSE is part of how I see myself. \\
\cline{2-2}
  & I have a lot in common with the average… and Members of these groups have a lot in common with each other. \\
\hline

\multirow{3}{5.8cm}{Autonomy} 
  & I decide how to do my job (method autonomy). \\
\cline{2-2}
  & I decide when to do tasks (scheduling autonomy). \\
\cline{2-2}
  & I can modify job objectives (criteria autonomy). \\
\hline

\multirow{6}{5.8cm}{Resilience – Monitoring} 
  & I’ve identified when my behaviour affected work. \\
\cline{2-2}
  & My team identified when our actions affected work. \\
\cline{2-2}
  & Project-level changes were identified. \\
\cline{2-2}
  & I identified external factors affecting my work. \\
\cline{2-2}
  & My team identified external work factors. \\
\cline{2-2}
  & External project-level factors were identified. \\
\hline

\multirow{2}{5.8cm}{Resilience – Responding} 
  & I responded to change using known techniques. \\
\cline{2-2}
  & I responded to change using new techniques. \\
\hline

\multirow{2}{5.8cm}{Resilience – Learning} 
  & I applied lessons from prior experience. \\
\cline{2-2}
  & I learned from superficially similar situations. \\
\hline

\multirow{2}{5.8cm}{Resilience – Anticipating} 
  & I predicted changes at work before they arose. \\
\cline{2-2}
  & I predicted my own behavioural changes. \\
\hline

\multirow{2}{5.8cm}{Professional Outcomes \\– Success} 
  & I usually receive authorship for contributions. \\
\cline{2-2}
  & I am listed as first/last author equally. \\
\hline

\multirow{3}{5.8cm}{Professional Outcomes \\– Satisfaction} 
  & I am paid fairly. \\
\cline{2-2}
  & I’m satisfied with salary increase chances. \\
\cline{2-2}
  & I’m satisfied with promotion chances. \\
\hline

\multirow{2}{5.8cm}{Professional Outcomes \\– Meaningfulness} 
  & I have found a meaningful career. \\
\cline{2-2}
  & My work serves a greater purpose. \\
\hline

\multirow{2}{5.8cm}{Team Communication} 
  & It’s easy to talk openly in my team. \\
\cline{2-2}
  & Team members frequently help one another. \\
\hline

\multirow{3}{5.8cm}{Attachment Style \\– Avoidant} 
  & I usually discuss my problems. (Reversed) \\
\cline{2-2}
  & I tell close partners everything. (Reversed) \\
\cline{2-2}
  & I don’t feel comfortable opening up. \\
\hline

\multirow{3}{5.8cm}{Attachment Style \\– Anxious} 
  & I worry others won’t care about me. \\
\cline{2-2}
  & I worry about losing close relationships. \\
\cline{2-2}
  & I get upset if a partner shows no interest. \\
\hline
\end{tabular}
\end{table}

\subsection{Data Analysis}
\label{DataAnalysis}
\subsubsection{Analysis of Social Media and Blog Posts}
\label{socialmediaandblogposts_analysis}
To examine the evolution of the RSE professional identity, social media and blog posts were analyzed to test Hypotheses 1-4, using the Linguistic Inquiry and Word Count Software (LIWC-22) \cite{Boyd2022} . 

A time period variable was created for the tweets and blog posts. Each was coded as belonging to one of three chronological periods that marked significant milestones in the development of the RSE identity. The first period began on January 1, 2012 and ran until the day before the first RSE Conference, September 14, 2016. In this period, the concept of the RSE was created and named.  The second period began with the first day of the first RSE Conference, September 15, 2016 \footnote{C. Woods, personal communication, January 17, 2020}. The second period ended the day before the fourth RSE Conference, September 16, 2019.  The third period began on September 17, 2019, the day of the fourth RSE Conference, and continued through the date of data collection, 8 September 2022. That conference featured an RSE Worldwide session that led to the formation of multiple international RSE groups, which grew in 2019-2020. During this period, the “first generation” RSEs had moved into senior and permanent positions and more junior RSEs\footnotemark[1] were employed.  We assessed whether the time period predicted the variation in three sets of LIWC-22 derived variables \cite{Boyd2017}: social identification, collective action, and linguistic style.  

Linguistic style was assessed following the methodology of Back et al. \cite{back2018we}. The LIWC scores of each dataset were z-transformed, and the absolute value of each z-score was taken. These absolute z-values were then summed to produce an overall LIWC score for each text. The average overall LIWC scores were then calculated for the group (such as blogposts). The deviation in linguistic style of each individual text from the average was calculated by subtracting the text’s score from the average score. 

A supplementary analysis was conducted using only LIWC-22 function word categories (prepositions, articles, auxiliary verbs, adverbs, conjunctions, personal and impersonal pronouns, negations), consistent with established Language Style Matching procedures.

Topic modelling was also performed to explore thematic structures in the texts, using LIWC’s Meaning Extraction Method (MEM) \cite{chung2008revealing} \cite{Boyd2017}. The MEM converted words to their base forms, retained 1–3 word phrases (N-grams), and generated a document–term matrix (DTM) based on the relative frequency of terms. Thresholds were set at $\ge5\%$ occurrence for blogs and $\ge5\%$ occurrence for tweets to account for text length differences. A Principal Components Analysis (PCA) with varimax rotation was then applied to the DTM to identify latent themes in the discourse.

\subsubsection{Analysis of Survey Data}
\label{analysisofsurveydata_analysis}
To adress RQ2, survey responses were examined to test Hypotheses 5-9, as described in Section \ref{sec:hypotheses} and Table \ref{tab:aims_hypotheses_analysis2}.  Analyses within R were conducted to generate descriptive statistics such as mean, percentages, and standard deviations, as well as inferential statistics such as linear regressions and ANOVAs to identify relationships among social identification, autonomy, resilience, and professional outcomes.  

Causal mediation analyses were performed using the \texttt{mediate} function from the \texttt{mediation} package in R \cite{tingley2014mediation} to test the indirect effects of social identification on professional outcomes via autonomy and resilience.  For all mediation analyses, we input the level of social identification as an RSE (measured as participants’ response to the single-item RSE social identification measure) as the independent variable, the overall autonomy (averaged across the three types of autonomy) or resilience score (averaged across the three levels of resilience and the four resilience abilities) as the mediator, and professional outcomes (success, satisfaction, meaningfulness, communication, and the average of the four outcomes) as the dependent variable. 

Linear regression was conducted to investigate the relationships of social identification as an RSE with professional outcomes, autonomy, and resilience. Linear regression was also used to investigate the relationship between these individual differences and both resilience and autonomy.  A one-way ANOVA was used to investigate the participants’ level of social identity, resilience, autonomy and professional outcomes.


\section{Results}

The results are presented in two subsections: \ref{CollectiveIdentity} presents the computational linguistic analyses, illustrating how the professional RSE identity evolved over time, and \ref{IndividualIdentification} provides the quantitative analysis used to predict resilience, autonomy, and professional outcomes.

\subsection{Social Identity Formation}
\label{CollectiveIdentity}

Variables associated with social identification, a focus on collective action, and linguistic style were measured using LIWC-22 categories \cite{Boyd2022}. We then analyzed whether the time period predicted each of these outcome variables. We ran linear regression models to investigate H1, H2, and H3. Topic modelling was conducted to investigate H4.  

\subsubsection{Twitter}
Analyses were run on the full tweet dataset as well as on the subset of this data containing the terms “RSE” or “research software” (e.g. “research software engineer,” “research software engineering”). Where no meaningful differences in results were found, only the results for the full dataset are reported.

Supporting our hypotheses H1a-H1c, the use of first-person singular pronouns decreased over time
, indicating a gradual reduction in self-focused language. In contrast, first-person plural 
and third-person plural pronouns 
increased, reflecting a growing emphasis on collective identity and references to others in the discourse. 


\begin{table}[h!]
\footnotesize
\centering
\caption{
Variables Over Time in Tweets}
\label{tab:social_id}
\begin{tabular}{p{2.2cm}rrrcl}
\textbf{Social Identification} & \textbf{R\textsuperscript{2}} & \textbf{F(df)} & \textbf{b} & \textbf{p} & \textbf{Trend} \\
\hline
First-person singular & 0.13 & -- & -0.01 & $<$.001 & Decreased \\
First-person plural & 0.01 & 252.4 (1, 28410) & 0.40 & $<$.0001 & Increased \\
Third-person plural & 0.00 & 121.6 (1, 28410) & 0.11 & $<$.0001 & Increased \\
\hline
\textbf{Collective Action} \\
\hline
Negative emotion & 0.00 & -- & -0.02 & $<$.001 & Decreased \\
Anger & 0.00 & 6.35 (1, 28410) & -0.01 & $<$.05 & Decreased \\
Discrepancy & 0.00 & 15.01 (1, 28410) & -0.07 & $<$.0001 & Decreased \\
Focus on the past & 0.00 & 31.45 (1, 28410) & 0.13 & $<$.0001 & Increased \\
Differentiation & 0.00 & 74.02 (1, 28410) & 0.19 & $<$.0001 & Increased \\
\hline
\multicolumn{2}{l}{\textbf{Supplementary Analysis}} \\
\hline
Third-person plural pronouns & 0.00 & 1.65 (1, 4283) & 0.04 & 0.20 & No change \\
Negative emotion & 0.01 & 0.23 (1, 4283) & 0.01 & 0.63 & No change \\
\hline
\textbf{Linguistic Style} \\
\hline
LIWC categories & 0.08 & 2362 (1, 28410) & -6.11 & $<$.0001 & Decreased \\
Functional words & 0.05 & 1342 (1, 28410) & -0.50 & $<$.0001 & Decreased \\
\hline
\end{tabular}

\textit{R$^2$ = coefficient of determination; F(df) = F statistic with degrees of freedom; b = regression coefficient; p = significance value; Trend = direction of change over time}.

\end{table}

As Table \ref{tab:social_id} shows, several collective action variables changed over time. Negative emotion decreased
, as did anger 
and discrepancy. 
In support of hypotheses H2a, H2b, and H2d, the changes indicate a reduction in expressions of negative affect and perceived gaps or inconsistencies in the discourse. By contrast, and countering hypotheses H2c and H2e, focus on the past increased
, as did differentiation
, suggesting a growing emphasis on temporal reflection and contrastive distinctions over time. 

Supplementary analysis of the subset revealed that most patterns observed in the full dataset were retained, with two exceptions
: use of third-person plural pronouns did not change significantly over time
, and negative emotion remained stable
.
Contrary to H3, linguistic style similarity decreased over time
. Using the full set of LIWC categories, deviation from the group average increased significantly
, and a similar pattern was observed when considering only functional words
. These findings indicate that individual posts became less aligned with the overall linguistic style over the observed period, suggesting increasing diversity in communication patterns across accounts.

Topic modeling revealed themes related to the RSE social identity and community activities. A principal components analysis of the document term matrix produced three factors: News, Discussion and Support, and Projects. The first factor, News, included references to news or newsletters and announcements of opportunities  (e.g., calls for grant applications, virtual sessions, programs). Discussion and Support contained references to teams, virtual drop-in sessions, membership, issues, questions, and chats. Terms loading on the Projects factor related to grants, applications, services, proposals, data storage, and expertise. 

Topic modeling of the subset dataset produced nearly identical themes, with the exception that Projects was the second (rather than the third), and that the third factor was Community (rather than Discussion and Support). Community included terms referring to particular events and associations, mentions of specific Twitter handles and web addresses, conferences, the international nature of what was being discussed, opportunities to sign up, and discussions. These results support hypothesis H4.

\subsubsection{Blog Posts} The first-person singular pronoun usage decreased over time
, indicating reduced self-focused language in the blog discourse (Table \ref{tab:social_id_blog}). In contrast, first-person plural pronouns increased
, reflecting greater emphasis on collective identity. Third-person plural pronoun usage did not change significantly
, suggesting that references to others outside the group remained stable. These results support hypotheses H1a and H1b, but not H1c.

The analysis supports four (H2a-H2c and H2e) of the five hypotheses related to collective action processes
. Negative emotion decreased over time
, as did anger
, focus on the past
, and differentiation
, indicating a reduction in negative affect and contrasting statements within blog posts. Discrepancy did not change significantly
, suggesting that perceived inconsistencies or gaps in the discourse remained stable. Hence, H2d was not supported.

The linguistic style similarity of blog posts decreased over time
. Using all LIWC categories, deviation from the group average increased significantly
. A similar pattern occurs when analyzing only functional words
. These results do not support H3.

As expected, the topics identified through topic modelling were related both to the RSE community and to its advancement. The principal components analysis of the document term matrix produced three factors labeled as: Contact/Staying in Touch, Shop Talk, and Community Progress. The first
included terms referring to discussions, posts, and contact between members. Shop Talk made reference to coding, testing, ease and difficulty, documentation, features, requirements, etc. Finally, Community Progress incorporated references related to social identification as well as collective action on behalf of the group, for example: software sustainability, community, funding, programs, goals, activities, funding, awareness, encouragement, and membership. These findings support H4.

\begin{table}[h!]
\footnotesize
\centering
\caption{
Variables Over Time in Blog Posts}
\label{tab:social_id_blog}
\begin{tabular}{lrrccl}
\textbf{Social Identification} 
& \textbf{R\textsuperscript{2}} & \textbf{F(df)} & \textbf{b} & \textbf{p} & \textbf{Trend} \\
\hline
First-person singular & 0.01 & 22.06 (1, 1762) & -0.17 & $<$.0001 & Decreased \\
First-person plural & 0.00 & 7.82 (1, 1762) & 0.09 & $<$.01 & Increased \\
Third-person plural & 0.00 & 0.86 (1, 1762) & -0.02 & 0.35 & No change \\

\hline
\textbf{Collective Action} \\ 
\hline
Negative emotion & 0.02 & 26.96 (1, 1762) & -0.03 & $<$.0001 & Decreased \\
Anger & 0.00 & 6.76 (1, 1762) & -0.01 & $<$.01 & Decreased \\
Focus on the past & 0.00 & 8.54 (1, 1762) & -0.14 & $<$.01 & Decreased \\
Differentiation & 0.01 & 14.42 (1, 1762) & -0.14 & $<$.001 & Decreased \\
Discrepancy & 0.00 & 1.18 (1, 1762) & -0.03 & 0.28 & No change \\

\hline
\textbf{Linguistic Style} \\ 
\hline
LIWC categories & 0.01 & 25.61 (1, 1763) & -2.75 & $<$.0001 & Decreased \\
Functional words & 0.01 & 10.98 (1, 1763) & -0.17 & $<$.001 & Decreased \\
\end{tabular}

\end{table}

\subsection{Orientation toward a Social Identity}
\label{IndividualIdentification}

\textbf{RSE work} Participants had been writing software for an average of $M = 8.99$ years ($SD = 1.04$), and the mean for engagement in RSE work was $M = 5.11$ (out of 7) ($SD = 1.04$, $\alpha = .72$), indicating that participants were indeed engaged in RSE work.
\\
\textbf{Social identity} 76.90\% of participants claimed the RSE identity, while 1.05\% of participants identified only with RSE and not with any academic discipline. A majority of those who did claim an additional disciplinary identity identified with Computer Science (35.17\%), Psychology (18.64\%), and Humanities (15.22\%). The level of social identification with the RSE or alternative identity measured as a single-item were not significantly different (Table \ref{tab:social_identification}).

\begin{table}[ht]
\centering
\footnotesize
\caption{Levels of identity, autonomy and outcomes}
\begin{tabular}{p{6cm}cc}
\textbf{Social Identity
} & \textbf{Mean} & \textbf{SD} \\
\hline
Single-item measure of social identification as an RSE & 5.64 & 1.50 \\
Single-item measure of other professional social identification & 5.80 & 1.59 \\
RSE self-investment & 5.49 & 1.59 \\
Other professional self-investment & 5.87 & 1.18 \\
RSE self-definition & 5.38 & 1.28 \\
Other professional self-definition & 5.67 & 1.28 \\
\hline
\textbf{Autonomy} \\
\hline
Overall Autonomy (mean of three types) & 5.60 & 0.96 \\
Method Autonomy & 5.71 & 1.19 \\
Scheduling Autonomy & 5.57 & 1.21 \\
Criteria Autonomy & 5.53 & 1.23 \\
\hline
\textbf{Professional Outcome} \\
\hline
Overall Outcomes (mean of three outcomes) & 5.36 & 0.79 \\
Success        & 5.30 & 1.10 \\
Satisfaction   & 4.88 & 1.45 \\
Meaningfulness & 5.48 & 1.14 \\
Team Communication & 5.56 & 1.07 \\
\end{tabular}
\label{tab:social_identification}
\end{table}

\textbf{Autonomy} Participants’ overall level of autonomy, averaged across the three types of autonomy, was 5.60 out of 7
. A one-way ANOVA revealed their levels of autonomy did not significantly differ across the types (F(2) = 2.15, p = .12).

\textbf{Professional outcomes} Participants’ overall level of positive professional outcomes (averaged across success, satisfaction, and meaningfulness) was 5.36 out of 7
. A one-way ANOVA demonstrated that the level of professional outcomes did significantly differ (F(2) = 21.04, p < .001). Tukey’s HSD Test for multiple comparisons found that participants’ level of satisfaction was significantly lower than their levels of success 
(p < .001, 95\% C.I. = [-0.65, -0.20]) and of meaningfulness 
(p < .001, 95\% C.I. = [0.28, 0.83]). Their levels of success and meaningfulness did not differ significantly from one another (p = .15). 

\textbf{Resilience} Participants’ overall level of resilience, averaged across the three levels – individual, team, and project – and four abilities, was 5.34 out of 7 (Table \ref{tab:resilience}). Averaging across the three levels, participants’ ability to anticipate 
was significantly lower than their ability to monitor 
(p = .02, 95\% C.I. = [-0.40, -0.02]) and to learn 
(p = .01, 95\% C.I. = [-0.43, -0.05]).

\begin{table}[h!]
\centering
\footnotesize
\caption{Level of resilience}
\resizebox{\columnwidth}{!}{%
\begin{tabular}{p{2cm}cccc}
\textbf{} & \textbf{Individual} & \textbf{Team} & \textbf{Project} & \textbf{Aggregate} \\
\midrule
Overall Resilience 
& 5.35 (SD = 0.85) & 5.25 (SD = 0.88) & 5.43 (SD = 0.88) & 5.34 (SD = 0.79) \\
Mean of Respond     & 5.36 & 5.26 & 5.37 & 5.33 (SD = 0.95) \\
Mean of Monitor     & 5.44 & 5.17 & 5.52 & 5.38 (SD = 0.92) \\
Mean of Learn       & 5.41 & 5.35 & 5.47 & 5.41 (SD = 0.91) \\
Mean of Anticipate  & 5.12 & 5.11 & 5.27 & 5.17 (SD = 1.09) \\
\end{tabular}
}
\label{tab:resilience}
\end{table}

\textbf{Social Identification as a Predictor of Professional Outcomes, Autonomy, and Resilience}
We ran linear regression models to investigate the relationships of social identification as an RSE (using the single-item measure of RSE social identification) with professional outcomes, autonomy, and resilience. As Table \ref{tab:si_regression} shows, social identification as an RSE predicted more positive professional outcomes in the aggregate (averaged across the four outcomes)
, as well as higher levels of autonomy 
and of resilience
. These results support H5–H7.

\begin{table}[ht]
\footnotesize
\centering
\caption{Regression Results: Social Identification as an RSE Predicting Autonomy, Resilience and Professional Outcomes}
\label{tab:si_regression}
\begin{tabular}{lcccc}
\textbf{Outcome} & \textbf{b} & \textbf{R\textsuperscript{2}} & \textbf{F(df)} & \textbf{p-value} \\
\hline
Professional Outcomes & 0.16 & 0.09 & 33.05 (1, 331) & $< .0001$ \\
Autonomy & 0.13 & 0.05 & 16.03 (1, 331) & $< .0001$ \\
Resilience & 0.23 & 0.19 & 78.82 (1, 328) & $< .0001$ \\
\end{tabular}
\end{table}

\textbf{Autonomy and Resilience as Mediators of the Relationship between Social Identification and Professional Outcomes}
We found that autonomy and resilience each mediated the relationship between social identification and professional outcomes. These analyses revealed significant mediation effects for each analysis.

In order to examine the potential mediation effects of autonomy and resilience, respectively, on the relationship between social identification and professional outcomes, we conducted 10 separate mediation analyses.   We conducted causal mediation analyses using the \texttt{mediate} function from the \texttt{mediation} package in R \cite{tingley2014mediation}. 

The positive relationship between social identification and averaged professional outcomes was partially mediated by overall autonomy (ACME = 0.07; p < .0001) (see Figure 1). The more participants embraced their RSE social identity, the more autonomous they felt, and the more autonomous they felt, the more positive professional outcomes they experienced. These results support H8a.

\begin{figure}[h] 
\centering
\includegraphics[trim=20 10 5 55, clip, scale=0.5]{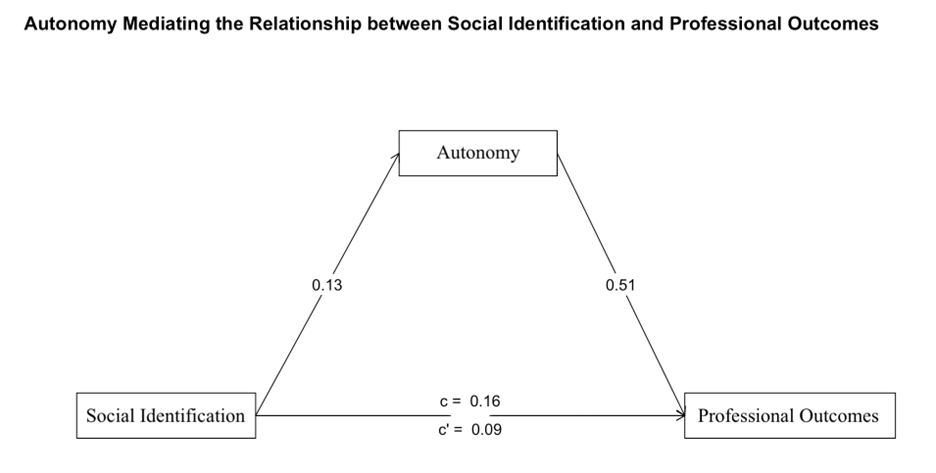}
\caption{Autonomy mediating the relationship between social identification and professional outcomes.}
\label{fig_rse1}
\end{figure}

Supporting H8b, positive relationship between social identification and averaged professional outcomes was fully mediated by overall resilience (ACME = 0.13; p < .0001) (see Figure 2). The more participants embraced their RSE social identity, the more resilient they were, and the more resilient they were, the more positive professional outcomes they experienced. 

\textbf{Individual Differences in Orienting Towards Versus Away from Threat as Predictors of Autonomy and Resilience}  
Social identification predicted higher levels of anxious attachment (R2 = 0.10, F(1, 330) = 35.54, p < .0001). Greater social identification as an RSE did not predict avoidant attachment (p = .73) or state or action orientation (p = .18). Secure attachment was negatively associated with social identification as an RSE (r = -0.25, p < .0001). These findings do not support H9a–H9c.

We investigated the relationship between these individual differences and both resilience and autonomy using linear regression models. In separate regressions, anxious attachment predicted significantly greater resilience (R2 = 0.23, F(1, 328) = 97.54, b = 0.27, p < .001), criteria autonomy (R2 = 0.03, F(1, 330) = 10.45, b = 0.15, p =.001) and, when adjusting for demographic variables (gender, age, income, education, and experience), method autonomy (R2 = 0.25, F(13, 311) = 7.89, b = 0.12, p = .01). State orientation also significantly predicted a higher level of method autonomy (R2 = 0.06, F(1, 330) = 20.62, b = 1.05, p < .0001). 
\begin{figure}[h] 
\centering
\includegraphics[trim=20 10 10 55, clip, scale=0.5]{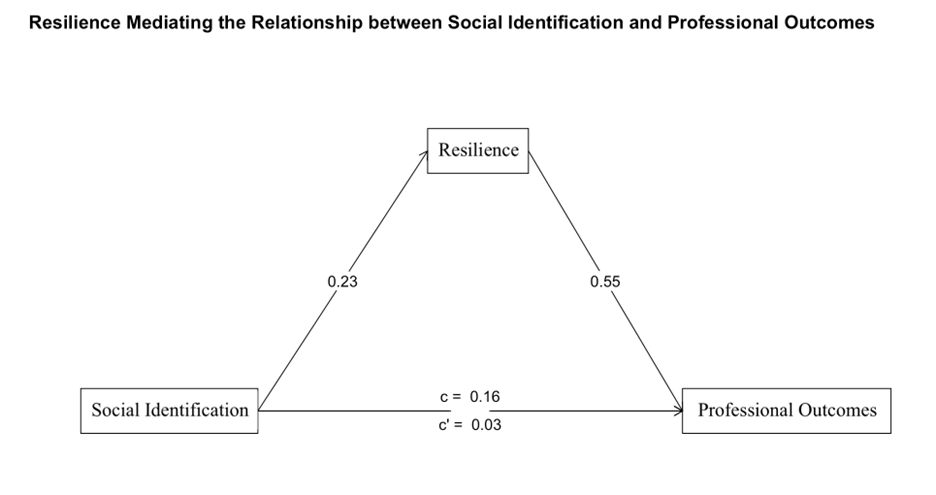}
\caption{Resilience mediating the relationship between social identification and professional outcomes.}
\label{fig_rse2}
\end{figure}
Avoidant attachment significantly predicted a lower level of criteria autonomy (R2 = 0.02, F(1, 329) = 6.14, b = -0.19, p = .01) and of learning (averaged across the three levels of resilience) (R2 = 0.02, F(1, 327) = 6.37, b = -.014, p = .01).
Secure attachment predicted lower levels of resilience  (R2 = 0.13, F(1, 328) = 49.28, b = -0.34, p < .0001) and did not significantly predict autonomy (p = .58).

\section{Discussion}
The following subsections discuss each research question in the context of our analysis of the RSE community, offer implications, and outline areas for future research.

\subsection{RQ1: How do social identities cohere over time?}
The analyses of public discourse indicate that an RSE identity emerged and cohered over the time studied. This is evidenced by support in the dataset of a decrease in singular first-person pronoun usage, and an increase in plural first-person pronoun usage. While use of the third-person plural only increased in the full Twitter dataset, the results align with an increase in social identification with the developing RSE identity. The analysis also reveals an identity for whom outgroup differentiation was not particularly central. This is consistent with the finding that 96\% of survey respondents claimed a disciplinary identity in addition to the RSE identity, because recognition and acceptance in the related research discipline is also important in this context \cite{Hovenden1996}.

Analysis also provided evidence that consolidation and recognition of the RSE identity over time may have accompanied gains for the professional group.  Anger decreased over time, as did negative emotion (there was only no significant change in negative emotion in the subset of tweets),  indicating relatedness and suggesting that the motivating emotions abated as the RSE identity crystallized and secured some of its intended effects for group members. 

The topic modelling also included expected themes related to the RSE social identity and to the advancement of the RSE community and its members. In both datasets (and the Twitter subset), themes included staying in touch with other members and sharing news, talk of work and projects, and providing one another and the community itself with support for success and advancement. These findings support the role of social identification in promoting the success, and recognition of the value, of the groups’ members.

\subsection{RQ2: How does orientation toward a social identity influence
professional outcomes?}
In the case of RSEs, greater social identification predicted more positive professional outcomes, and higher levels of autonomy and resilience.  In turn, autonomy and resilience mediated the relationship between social identification and professional outcomes. In line with existing theory and research, these findings suggest that a stronger sense of one’s identity, in this case professional, bolsters one’s ability to operate independently and overcome challenges.

Results also suggest that greater attention to threats may be associated with resilience. Social identification was associated with attachment anxiety (but not security), and attachment anxiety was positively associated with both resilience and with criteria and method autonomy. A state orientation likewise predicted greater method autonomy. In contrast, avoidant attachment was associated with feeling one had less autonomy in the choice of the goals one was supposed to achieve (criteria autonomy), and with a lower level of learning (one of Hollnagel's four resilience abilities \cite{Hollnagel2011}).

While secure attachment is typically found to be associated with more positive life outcomes, the findings here align with an understanding of attachment anxiety as a coercive strategy \cite{Crittenden2011} that involves an under-appreciation of one’s own power and efficacy. Attachment anxiety is associated with a lack of self-esteem and a low level of regard for oneself \cite{Bartholomew1991} and can be countered by strengthening an individual’s commitment to one’s goals and sense of self-efficacy  \cite{Arriaga2018}. These individuals are likely to be more in control than they relationally let on, in order to justify and maintain arousal around threats and potential threats \cite{Mikulincer2003a}. They may be in a better position to respond to threats and to learn from them. By contrast, a tendency to avoid threats undermines one’s ability to respond, and the ability to learn from experiences or to safeguard against them.
\subsection{Implications}
The details of the research reported here are specific to the case of RSEs. However, the study demonstrates that examining social identity in software engineering is both viable and worthwhile. 

Changes in technology and the organization of work are redefining SE roles and tasks.  AI tooling is increasingly being used to support \cite{Rauf2025} or replace development activities\cite{chen_ai_2026}. 
Understanding the social identity of SE groups provides a complementary approach to other developer-centred research and can help engineering groups manage the changing professional landscape.  
For example, Trinkenreich et al \cite{trinkenreichvirtual2022} investigated what factors contribute to a sense of belonging in open source communities. They found that social motives like maximising joint or others’ gains, kinship and altruism contribute to the formation of a sense of belonging. Similar factors featured in our and related social identity research \cite{Amiot2010}, including emotional change, community coherence, and relatedness. 

This analysis focused on professional outcomes that form part of a developer’s experience. Other outcomes are also important in software practice, including productivity \cite{Storey2022}, turnover \cite{kuutila2025staying} and career progression \cite{hochstein_developing_2023}. While we did not investigate the impact social identity may have on these outcomes, developer wellbeing underpins several of them, and social identity is known to have a positive effect on mental health and wellbeing {\cite{Haslam2018}. 

Similarly, the focus in this work on autonomy and threat may lend insight to related work. Autonomy is central to agility at team and organisational level \cite{lassenius2025} and of self-organising teams \cite{Hoda2013}. However, the quality is difficult to assess and remains a lively focus of debate in research \cite{acharya2019} and practitioner circles \cite{Dingsøyr2022}.  Developers that question poor design \cite{lopez2023accounting} or `speak up' about insufficient security \cite{lopez2023security} have been shown to give responses that support resilient performance and can improve software. Likewise, the ability to adapt performance under stress is also an aspect of performance in systems where developers diagnose and fix run-time failures \cite{cook2020above}.  

\subsection{Future Work} 
\paragraph{To map the social identity of RSEs.} In the analysis of social media and blog posts, the pattern of findings for some collective action-related variables was mixed. Discrepancy decreased over time in the Twitter dataset, but had a negative nonsignificant relationship in the blog dataset. Focus on the past, and differentiation decreased in the blog dataset, but increased in the Twitter dataset. Findings surrounding the past, discrepancy, and differentiation invite further research using different, more extensive datasets. 

Counter to our hypothesis, the similarity of posts’ linguistic style decreased over time. It may be that as a social group becomes better established and the “we” of the identity is increasingly embraced, individual group members become freer to express themselves differently, and discuss a wider variety of identity-relevant themes. Further research should explore the contexts in which RSEs and other groups are likely to share linguistic style.

The relatively high level of autonomy reported by survey participants was surprising given anecdotal complaints gathered about RSEs’ perceived level of autonomy, a discrepancy that warrants further investigation. Likewise, rationalization (minimizing and attending less to a threat) is more likely to occur when responses to threat are expected not to make a difference \cite{Laurin2012}, which aligns with survey results that avoidant individuals are less likely to feel in control of the larger circumstances in which they are operating.  Curiously, secure attachment was negatively associated with resilience and may indicate naive, rather than earned secure attachment. In such cases, secure attachment can develop when people have not faced major threats or dangers, and have instead grown up in a safe and comfortable environment. Future research should examine this result and verify that the positive association of anxious (rather than secure) attachment with resilience and autonomy are not artifacts of self-report and social desirability bias.

\paragraph{In software engineering} The research design used here combines quantitative analysis with self-report that can be adapted and applied to identify social identity in other SE groups. An extended analysis should assess how findings in other contexts, such as developer experience in companies \cite{greiler_actionable_2022} or open-source projects \cite{trinkenreich_pots_2021}, align or extend findings associated here with social identity.

\paragraph{In RSE practice}
Efforts  are currently underway to recognize \textit{all} those who contribute directly to research projects, including those who run facilities or computational platforms. Suggested names include "Research Technical Professional", and "Digital Research Technical Professional".  Anecdotally, many professionals prefer labels that have a historical context and precisely describe what they do, such as "infrastructure engineer" or "experimental officer". Time and additional research will clarify which names are accepted, and whether they represent distinct social identity groups.

\subsection{Threats to validity} 

\paragraph{External validity}
Due to the linguistic analysis, this study is by necessity constrained to
those RSEs who feel comfortable expressing themselves publicly in English.
Although we have sourced texts from around the world, and sent the survey widely,
the majority of samples are from Europe, notably the UK (Tables 
\ref{tab:twitter_accounts} and \ref{tab:blogs_overview}).
This is to be expected, as there will be more engagement and available data where the social identity started.
To mitigate this, other national and continental RSE communities must be studied.

\paragraph{Construct validity}
An analysis of individual words or n-grams cannot fully capture the richness of
discourse of a community and may misrepresent non-native speakers.
This may partly explain the lack of significance of some results and could be mitigated through additional analysis of the dataset with more sophisticated natural language processing. 
The effect sizes for the significant findings reported were in the small to moderate range, indicating that the predictors discussed should be interpreted as important but not dominant contributors among others. This is typical in social and organizational psychology, and does not imply triviality. Future work should examine how the predictors discussed here interact with additional factors to better account for variability in professional outcomes.

\paragraph{Internal validity}
Due to the method followed, we don't know to what extent 
the authors of tweets and blogs overlap with the survey's respondents. 
The two parts of the study are likely to have different populations, so our results cannot shed light on whether an 
individual's evolution of their own RSE identity influences autonomy, resilience and professional outcomes. Future research, with a different method, would be needed to investigate this.

\section{Conclusions}

This paper provides evidence for the evolution of a RSE social identity, and evidence that orienting towards that identity has had positive effects on RSEs in terms of autonomy, resilience and professional outcomes. These are important results for the management and wellbeing of RSEs, but also for other software engineering communities who may benefit from reflecting on their own group identities. With its theoretical grounding in social psychology, this work foregrounds a new approach for research in the social aspects of software engineering.  Today, communities of developers combine and recombine in professional contexts.  This study illustrates a way to systematically identify professional groups, and to understand factors that affect individual practice.

\begin{acks}

We thank Ryan Boyd, Tina Keil and Charlotte Entwistle for their work in collecting and analysing the data. 
This research was supported by EPSRC grants EP/T017465/1, EP/T016779/1 and EP/T017198/1 for STRIDE (Socio-technical resilience in software development). 

\end{acks}

\section*{Supplementary Materials}
\label{sec:supplementary-materials}
Supplementary materials for this study are available at: \url{https://doi.org/10.21954/ou.rd.31681381}. Open registration documents for blogs: \url{https://doi.org/10.17605/OSF.IO/N54AY}, Twitter: \url{https://doi.org/10.17605/OSF.IO/82W7K}, and the survey: \url{https://doi.org/10.17605/OSF.IO/6ZGPT}

\printbibliography

@article{chen_ai_2026,
	title = {{AI} is threatening science jobs. {Which} ones are most at risk?},
	volume = {651},
	copyright = {2026 Springer Nature Limited},
	issn = {1476-4687},
	doi = {10.1038/d41586-026-00444-9},
	language = {en},
	number = {8104},
	urldate = {2026-04-28},
	journal = {Nature},
	publisher = {Nature Publishing Group},
	author = {Chen, Edward},
	month = feb,
	year = {2026},
	pages = {19--20}
}

@book{robson2016real,
  title={Real world research},
  author={Robson, Colin and McCartan, Kieran},
  year={2016},
edition={4th},
publisher={John Wiley \& Sons}
}

@article{gustafsson2014selection,
  title={Selection bias in choice of words: Evaluations of “I” and “we” differ between contexts, but “they” are always worse},
  author={Gustafsson Send{\'e}n, Marie and Lindholm, Torun and Sikstr{\"o}m, Sverker},
  journal={Journal of Language and Social Psychology},
  volume={33},
  number={1},
  pages={49--67},
  year={2014},
  publisher={Sage Publications Sage CA: Los Angeles, CA}
}

@article{koschate2021asia,
  title={ASIA: Automated social identity assessment using linguistic style},
  author={Koschate, Miriam and Naserian, Elahe and Dickens, Luke and Stuart, Avelie and Russo, Alessandra and Levine, Mark},
  journal={Behavior Research Methods},
  volume={53},
  number={4},
  pages={1762--1781},
  year={2021},
  publisher={Springer}
}

@article{sharp2002software,
  title={Software engineering: community and culture},
  author={Sharp, Helen and Robinson, Hugh and Woodman, Mark},
  journal={IEEE Software},
  volume={17},
  number={1},
  pages={40--47},
  year={2002},
  publisher={IEEE}
}

@article{kuutila2025staying,
  title={Staying or Leaving? How Job Satisfaction, Embeddedness and Antecedents Predict Turnover Intentions of Software Professionals},
  author={Kuutila, Miikka and Ralph, Paul and Qiu, Huilian Sophie and Santos, Ronnie de Souza and Choetkiertikul, Morakot and Alkadhi, Rana and Devroey, Xavier and Robles, Gregorio and Hata, Hideaki and Baltes, Sebastian and others},
  journal={arXiv preprint arXiv:2512.00869},
  year={2025}
}

@inproceedings{fagerholm_developer_2012,
	title = {Developer experience: {Concept} and definition},
	shorttitle = {Developer experience},
	OPTurl = {https://ieeexplore.ieee.org/abstract/document/6225984/},
	OPTurldate = {2023-11-04},
	booktitle = {Int'l Conf. on software and system process},
	publisher = {IEEE},
	author = {Fagerholm, Fabian and Münch, Jürgen},
	year = {2012},
	pages = {73--77},
}

@article{greiler_actionable_2022,
	title = {An actionable framework for understanding and improving developer experience},
	volume = {49},
	OPTurl = {https://ieeexplore.ieee.org/abstract/document/9785882/},
	number = {4},
	OPTurldate = {2023-11-04},
	journal = {IEEE Trans. on Software Engineering},
	author = {Greiler, Michaela and Storey, Margaret-Anne and Noda, Abi},
	year = {2022},
	OPTnote = {Publisher: IEEE},
	pages = {1411--1425},
}

@article{sharp_models_2009,
	title = {Models of motivation in software engineering},
	volume = {51},
	OPTurl = {https://www.sciencedirect.com/science/article/pii/S0950584908000827?casa_token=UGXA2rgTz5cAAAAA:qoQgyASHkJMqpzezaDDH9efsZkOIzeAXYLMJXoH5ubB-_RJmbQmZBPEtzCR5N-zp5UTZDS4x9w},
	number = {1},
	OPTurldate = {2023-11-04},
	journal = {Information and software technology},
	author = {Sharp, Helen and Baddoo, Nathan and Beecham, Sarah and Hall, Tracy and Robinson, Hugh},
	year = {2009},
	OPTnote = {Publisher: Elsevier},
	pages = {219--233},
}

@article{trinkenreich_pots_2021,
	title = {Pots of {Gold} at the {End} of the {Rainbow}: {What} is {Success} for {Open} {Source} {Contributors}?},
	volume = {48},
	shorttitle = {Pots of {Gold} at the {End} of the {Rainbow}},
	OPTurl = {https://ieeexplore.ieee.org/abstract/document/9524493/},
	number = {10},
	OPTurldate = {2023-11-04},
	journal = {IEEE Trans. on Software Engineering},
	author = {Trinkenreich, Bianca and Guizani, Mariam and Wiese, Igor and Conte, Tayana and Gerosa, Marco and Sarma, Anita and Steinmacher, Igor},
	year = {2021},
	OPTnote = {Publisher: IEEE},
	pages = {3940--3953},
}

@inproceedings{trinkenreich_unraveling_2024,
	OPTaddress ={Lisbon Portugal},
	title = {Unraveling the {Drivers} of {Sense} of {Belonging} in {Software} {Delivery} {Teams}: {Insights} from a {Large}-{Scale} {Survey}},
	OPTisbn = {9798400702174},
	shorttitle = {Unraveling the {Drivers} of {Sense} of {Belonging} in {Software} {Delivery} {Teams}},
	OPTurl = {https://dl.acm.org/doi/10.1145/3597503.3639119},
	OPTdoi = {10.1145/3597503.3639119},
	language = {en},
	OPTurldate = {2024-05-02},
	booktitle = {46th ICSE},
	publisher = {ACM},
	author = {Trinkenreich, Bianca and Gerosa, Marco Aurelio and Steinmacher, Igor},
	OPTmonth = apr,
	year = {2024},
	pages = {1--12},
}

@article{moe_attractive_2023,
	title = {Attractive {Workplaces}: {What} are {Engineers} {Looking} {For}?},
	shorttitle = {Attractive {Workplaces}},
	OPTurl = {https://ieeexplore.ieee.org/abstract/document/10128875/},
	OPTurldate = {2024-05-02},
	journal = {IEEE Software},
	author = {Moe, Nils Brede and Stray, Viktoria and Smite, Darja and Mikalsen, Marius},
	year = {2023},
	OPTnote = {Publisher: IEEE},
}

@article{hochstein_developing_2023,
	title = {Developing {Your} {Software} {Engineering} {Career}: {Words} of {Advice} {From} {Seasoned} {Professionals}},
	volume = {40},
	shorttitle = {Developing {Your} {Software} {Engineering} {Career}},
	OPTurl = {https://www.computer.org/csdl/api/v1/periodical/mags/so/2023/05/10273788/1R6sOd7L64o/download-article/pdf},
	number = {05},
	OPTurldate = {2024-05-02},
	journal = {IEEE Software},
	author = {Hochstein, Lorin and Lanubile, Filippo and Nolan, Laura and Prikladnicki, Rafael},
	year = {2023},
	OPTnote = {Publisher: IEEE Computer Society},
	pages = {29--33},
}

@article{lopez2023security,
  title={Security responses in software development},
  author={Lopez, Tamara and Sharp, Helen and Bandara, Arosha and Tun, Thein and Levine, Mark and Nuseibeh, Bashar},
  journal={ACM Trans. on Software Engineering and Methodology},
  volume={32},
  number={3},
  pages={1--29},
  year={2023},
  publisher={ACM New York, NY, USA}
}

@inproceedings{lopez2023accounting,
  title={Accounting for socio-technical resilience in software engineering},
  author={Lopez, Tamara and Sharp, Helen and Wermelinger, Michel and Langer, Melanie and Levine, Mark and Jay, Caroline and Yu, Yijun and Nuseibeh, Bashar},
  booktitle={16th Int'l Conf. on Cooperative and Human Aspects of Software Engineering},
  pages={31--36},
  year={2023},
  organization={IEEE}
}

@article{cook2020above,
  title={Above the line, below the line},
  author={Cook, Richard I},
  journal={Communications of the ACM},
  volume={63},
  number={3},
  pages={43--46},
  year={2020},
  publisher={ACM New York, NY, USA}
}

@INPROCEEDINGS{Baeckevik2019,
  author={Bäckevik, Andreas and Tholén, Erik and Gren, Lucas},
  booktitle={12th Int'l Workshop on Cooperative and Human Aspects of Software Engineering}, 
  title={Social Identity in Software Development}, 
  year={2019},
  pages={107-114},
  publisher = {IEEE},
  OPTdoi={10.1109/CHASE.2019.00033}
}

@article{Parr2013,
    author = {Parr, Chris},
	title = {Save your work--give software engineers a career track},
	volume = {15},
	journal = {Times Higher Education},
	year = {2013},
  OPThowpublished = {\url{https://www.timeshighereducation.com/news/save-your-work-give-software-engineers-a-career-track/2006431.article}},
  OPTnote = {Accessed 2025-06-25} % Added access date
}

@article{Amiot2010,
	title = {Changes in social identities over time: {The} role of coping and adaptation processes},
	volume = {49},
	copyright = {http://onlinelibrary.wiley.com/termsAndConditions\#vor},
	issn = {0144-6665, 2044-8309},
	shorttitle = {Changes in social identities over time},
	OPTurl = {https://bpspsychub.onlinelibrary.wiley.com/doi/10.1348/014466609X480624},
	OPTdoi = {10.1348/014466609X480624},
	language = {en},
	number = {4},
	OPTurldate = {2025-09-01},
	journal = {British J. of Social Psychology},
	author = {Amiot, Catherine E. and Terry, Deborah J. and Wirawan, Dian. and Grice, Tim A.},
	month = nov,
	year = {2010},
	pages = {803--826},
}

@article{Smith_social_2015,
	title = {Social identity formation during the emergence of the occupy movement},
	volume = {45},
	copyright = {http://onlinelibrary.wiley.com/termsAndConditions\#vor},
	issn = {0046-2772, 1099-0992},
	OPTurl = {https://onlinelibrary.wiley.com/doi/10.1002/ejsp.2150},
	OPTdoi = {10.1002/ejsp.2150},
	language = {en},
	number = {7},
	OPTurldate = {2025-09-02},
	journal = {European J. of Social Psychology},
	author = {Smith, Laura G. E. and Gavin, Jeffrey and Sharp, Elise},
	month = dec,
	year = {2015},
	pages = {818--832},
}

@article{SmithThomasMcGarty2015,
	title = {“{We} {Must} {Be} the {Change} {We} {Want} to {See} in the {World}”: {Integrating} {Norms} and {Identities} through {Social} {Interaction}},
	volume = {36},
	issn = {0162-895X, 1467-9221},
	shorttitle = {“{We} {Must} {Be} the {Change} {We} {Want} to {See} in the {World}”},
	OPTurl = {https://onlinelibrary.wiley.com/doi/10.1111/pops.12180},
	OPTdoi = {10.1111/pops.12180},
	language = {en},
	number = {5},
	OPTurldate = {2025-09-02},
	journal = {Political Psychology},
	author = {Smith, Laura G. E. and Thomas, Emma F. and McGarty, Craig},
	month = oct,
	year = {2015},
	pages = {543--557},
}

@article{Langer2019,
	title = {Digital dissent: {An} analysis of the motivational contents of tweets from an {Occupy} {Wall} {Street} demonstration.},
	volume = {5},
	shorttitle = {Digital dissent},
	OPTurl = {https://psycnet.apa.org/record/2018-03958-001},
	number = {1},
	OPTurldate = {2025-09-08},
	journal = {Motivation Science},
	author = {Langer, Melanie and Jost, John T. and Bonneau, Richard and Metzger, Megan MacDuffee and Noorbaloochi, Sharareh and Penfold-Brown, Duncan},
	year = {2019},
	OPTnote = {Publisher: Educational Publishing Foundation},
	pages = {14},
	
}

@book{Ainsworth1978,
  author = {Ainsworth, Mary D. Salter and Blehar, Mary C. and Waters, Everett and Wall, Sally},
  year = {1978},
  title = {Patterns of attachment: A psychological study of the strange situation},
  publisher = {Lawrence Erlbaum}
}

@article{Arnold2007,
  author = {Arnold, Kathleen A. and Turner, Nick and Barling, Julian and Kelloway, E. Kevin and McKee, M. Catherine},
  year = {2007},
  title = {Transformational leadership and psychological well-being: the mediating role of meaningful work},
  journal = {J. of occupational health psychology},
  volume = {12},
  number = {3},
  pages = {193}
}

@article{Arriaga2018,
  author = {Arriaga, Ximena B. and Kumashiro, Mami and Simpson, Jeffry A. and Overall, Nickola C.},
  year = {2018},
  title = {Revising working models across time: Relationship situations that enhance attachment security},
  journal = {Personality and Social Psychology Review},
  volume = {22},
  number = {1},
  pages = {71-96}
}

@article{Bartholomew1991,
  author = {Bartholomew, Kim and Horowitz, Leonard M.},
  year = {1991},
  title = {Attachment styles among young adults: a test of a four-category model},
  journal = {J. of personality and social psychology},
  volume = {61},
  number = {2},
  pages = {226}
}

@article{Blader2017,
  author = {Blader, Steven L. and Yu, Serena},
  year = {2017},
  title = {Are status and respect different or two sides of the same coin?},
  journal = {Academy of Management Annals},
  volume = {11},
  number = {2},
  pages = {800-824}
}

@article{Bledow2021,
  author = {Bledow, Ronald and Kühnel, Jonas and Jin, Michael and Kuhl, Julius},
  year = {2021},
  title = {Breaking the chains: The inverted-U-shaped relationship between action-state orientation and creativity under low job autonomy},
  journal = {J. of Management},
  volume = {48},
  number = {4},
  pages = {905-935}
}

@book{Bowlby1969,
  author = {Bowlby, John},
  year = {1969},
  title = {Attachment and loss, Vol. I: Attachment},
  publisher = {Basic Books},
  OPTaddress ={New York},
  note = {(Reprinted 1982)}
}

@article{Boyes2020,
  author = {Boyes, Alice},
  year = {2020},
  title = {How to overcome your fear of making mistakes},
  journal = {Harvard Business Review}
}

@article{Brooke1988,
  author = {Brooke, Paul P. and Russell, Daniel W. and Price, James L.},
  year = {1988},
  title = {Discriminant validation of measures of job satisfaction, job involvement, and organizational commitment},
  journal = {J. of applied psychology},
  volume = {73},
  number = {2},
  pages = {139}
}

@article{Caldwell2012,
  author = {Caldwell, Jessica G. and Shaver, Phillip R.},
  year = {2012},
  title = {Exploring the cognitive-emotional pathways between adult attachment and ego-resiliency},
  journal = {Individual Differences Research},
  volume = {10},
  number = {3}
}

@article{Cano2004,
  author = {Cano, Jamie and Castillo, Julia X.},
  year = {2004},
  title = {Factors explaining job satisfaction among faculty},
  journal = {J. of Agricultural education},
  volume = {45},
  number = {3},
  pages = {65-74}
}

@misc{RSESociety2021,
  author = {{Society of Research Software Engineering}},
  year = {2021},
  title = {Celebrating 10 Years of the RSE movement!},
  howpublished = {\url{https://society-rse.org/events/celebrating-10-years-of-the-rse-movement/}},
  note = {Accessed 2025-06-25} % Added access date as it's a website
}

@book{Crittenden2011,
  author = {Crittenden, Patricia M. and Landini, Andrea},
  year = {2011},
  title = {Assessing adult attachment: A dynamic-maturational approach to discourse analysis},
  publisher = {WW Norton \& Co}
}

@article{Davis2017,
  author = {Davis, William A. and Jones, Stuart and Crowell-Kuhnberg, Ann M. and O'Keeffe, Devin and Boyle, Kelly M. and Klainer, Stephen B. and Yule, Steven},
  year = {2017},
  title = {Operative team communication during simulated emergencies: Too busy to respond?},
  journal = {Surgery},
  volume = {161},
  number = {5},
  pages = {1348-1356}
}

@article{Diefendorff2000,
  author = {Diefendorff, James M. and Hall, Rebecca J. and Lord, Robert G. and Strean, Matthew L.},
  year = {2000},
  title = {Action--state orientation: Construct validity of a revised measure and its relationship to work-related variables},
  journal = {J. of Applied Psychology},
  volume = {85},
  number = {2},
  pages = {250}
}

@article{Doosje1998,
  author = {Doosje, Bertjan and Branscombe, Nyla R. and Spears, Russell and Manstead, Antony S. R.},
  year = {1998},
  title = {Guilty by association: When one's group has a negative history},
  journal = {J. of personality and social psychology},
  volume = {75},
  number = {4},
  pages = {872}
}

@article{Downs1957,
  author = {Downs, Anthony},
  year = {1957},
  title = {An economic theory of political action in a democracy},
  journal = {J. of political economy},
  volume = {65},
  number = {2},
  pages = {135-150}
}

@article{Ghaferi2016,
  author = {Ghaferi, Amir A. and Dimick, Justin B.},
  year = {2016},
  title = {Importance of teamwork, communication and culture on failure-to-rescue in the elderly},
  journal = {J. of British Surgery},
  volume = {103},
  number = {2},
  pages = {e47-e51}
}

@article{Harpaz2002,
  author = {Harpaz, Itzhak and Fu, Xiaocong},
  year = {2002},
  title = {The structure of the meaning of work: A relative stability amidst change},
  journal = {Human relations},
  volume = {55},
  number = {6},
  pages = {639-667}
}

@book{Haslam2004,
  author = {Haslam, S. Alexander},
  year = {2004},
  title = {Social identity in organizations: The social identity approach},
  publisher = {Sage},
  OPTaddress ={London, UK}
}

@article{Haslam2016,
  author = {Haslam, Catherine and Cruwys, Tegan and Haslam, S. Alexander and Dingle, Genevieve and Chang, Mei X. L.},
  year = {2016},
  title = {Groups 4 Health: Evidence that a social-identity intervention that builds and strengthens social group membership improves mental health},
  journal = {J. of Affective Disorders},
  volume = {194},
  pages = {188-195}
}

@article{Haslam2008,
  author = {Haslam, Catherine and Holme, Alice and Haslam, S. Alexander and Iyer, Anjana and Jetten, Jolanda and Williams, W. H.},
  year = {2008},
  title = {Maintaining group memberships: Social identity continuity predicts well-being after stroke},
  journal = {Neuropsychological Rehabilitation},
  volume = {18},
  pages = {671-691}
}

@book{Haslam2018,
  author = {Haslam, Catherine and Jetten, Jolanda and Cruwys, Tegan and Dingle, Genevieve A. and Haslam, S. Alexander},
  year = {2018},
  title = {The new psychology of health: Unlocking the social cure},
  publisher = {Routledge},
  OPTaddress ={New York}
}

@article{Haslam1999,
  author = {Haslam, S. Alexander and Oakes, Penelope J. and Reynolds, Katherine J. and Turner, John C.},
  year = {1999},
  title = {Social identity salience and the emergence of stereotype consensus},
  journal = {Personality and social psychology bulletin},
  volume = {25},
  number = {7},
  pages = {809-818}
}

@misc{Hettrick2016,
  author = {Hettrick, Simon},
  year = {2016},
  title = {A not-so-brief history of Research Software Engineers},
  howpublished = {\url{https://www.software.ac.uk/blog/2016-08-17-not-so-brief-history-research-software-engineers-0}},
  note = {Accessed 2025-06-25} % Added access date
}

@article{Hogg2006,
  author = {Hogg, Michael A. and Reid, Scott A.},
  year = {2006},
  title = {Social identity, self-categorization, and the communication of group norms},
  journal = {Communication Theory},
  volume = {16},
  pages = {7-30}
}

@article{Hollnagel2014,
  author = {Hollnagel, Erik},
  year = {2014},
  title = {Resilience engineering and the built environment},
  journal = {Building Research \& Information},
  volume = {42},
  number = {2},
  pages = {221-228}
}

@book{Hollnagel2011,
  author = {Hollnagel, Erik},
  year = {2011},
  title = {Resilience Engineering and Safety Management},
  publisher = {Mines Paris Tech},
  OPTaddress ={Paris, France}
}

@incollection{Jetten2012,
  author = {Jetten, Jolanda and Haslam, S. Alexander and Haslam, Catherine},
  year = {2012},
  title = {The case for a social identity analysis of health and well-being},
  booktitle = {The social cure: Identity, health and well-being},
  OPTeditor ={Jetten, Jolanda and Haslam, Catherine and Haslam, S. Alexander},
  pages = {3-20},
  publisher = {Psychology Press},
  OPTaddress ={Hove}
}

@article{Jost2012,
  author = {Jost, John T. and Chaikalis-Petritsis, Vasso and Abrams, Dominic and Sidanius, Jim and Van Der Toorn, Jojanneke and Bratt, Christopher},
  year = {2012},
  title = {Why men (and women) do and don’t rebel: Effects of system justification on willingness to protest},
  journal = {Personality and social psychology Bulletin},
  volume = {38},
  number = {2},
  pages = {197-208}
}

@book{Kelly1996,
  author = {Kelly, Caroline and Breinlinger, Sara},
  year = {1996},
  title = {The social psychology of collective action: Identity, injustice and gender},
  publisher = {Taylor \& Francis US}
}

@article{Klohnen1998,
  author = {Klohnen, Eva C. and Bera, Sara},
  year = {1998},
  title = {Behavioral and experiential patterns of avoidantly and securely attached women across adulthood: a 31-year longitudinal perspective},
  journal = {J. of personality and social psychology},
  volume = {74},
  number = {1},
  pages = {211}
}

@incollection{Kuhl2000,
  author = {Kuhl, Julius},
  year = {2000},
  title = {A functional-design approach to motivation and self-regulation: The dynamics of personality systems interactions},
  booktitle = {Handbook of self-regulation},
  OPTeditor ={Boekaerts, Monique and Pintrich, Paul R. and Zeidner, Moshe},
  pages = {111-169},
  publisher = {Academic Press},
  OPTaddress ={San Diego, CA}
}

@inproceedings{uwasomba2016managing,
  title={Managing knowledge flows in Mauritian multinational corporations: Empirical analysis using the SECI model},
  author={Uwasomba, Chukwudi Festus and Seeam, Preetila and Bellekens, Xavier and Seeam, Amar},
  booktitle={2016 IEEE International Conference on Emerging Technologies and Innovative Business Practices for the Transformation of Societies (EmergiTech)},
  pages={341--344},
  year={2016},
  organization={IEEE}
}

@article{Laurin2012,
  author = {Laurin, Kelly and Kay, Aaron C. and Fitzsimons, Gail J.},
  year = {2012},
  title = {Reactance versus rationalization: Divergent responses to policies that constrain freedom},
  journal = {Psychological Science},
  volume = {23},
  number = {2},
  pages = {205-209}
}

@article{Manning2003,
  author = {Manning, T. T.},
  year = {2003},
  title = {Leadership across cultures: Attachment style influences},
  journal = {J. of Leadership \& Organizational Studies},
  volume = {9},
  number = {3},
  pages = {20-30}
}

@article{Mikulincer1997,
  author = {Mikulincer, Mario},
  year = {1997},
  title = {Adult attachment style and information processing: individual differences in curiosity and cognitive closure},
  journal = {J. of personality and social psychology},
  volume = {72},
  number = {5},
  pages = {1217}
}

@book{Mikulincer2003a,
  author = {Mikulincer, Mario and Shaver, Phillip R.},
  year = {2003},
  title = {The attachment behavioral system in adulthood: Activation, psychodynamics, and interpersonal processes}
}

@article{Mikulincer2003b,
  author = {Mikulincer, Mario and Shaver, Phillip R. and Pereg, Dalit},
  year = {2003},
  title = {Attachment theory and affect regulation: The dynamics, development, and cognitive consequences of attachment-related strategies},
  journal = {Motivation and emotion},
  volume = {27},
  number = {2},
  pages = {77-102}
}

@article{Muller1986,
  author = {Muller, Edward N. and Opp, Karl-Dieter},
  year = {1986},
  title = {Rational choice and rebellious collective action},
  journal = {American Political Science Review},
  volume = {80},
  number = {2},
  pages = {471-487}
}

@misc{Nord1990,
  author = {Nord, Walter R. and Brief, Arthur P. and Atieh, Judith M. and Doherty, Elizabeth M.},
  year = {1990},
  title = {Studying meanings of work: The case of work values}
}

@misc{OCarroll2021,
  author = {O'Carroll, Elizabeth},
  year = {2021},
  month = jun,
  title = {Building a career path for research software engineers},
  howpublished = {\url{https://research.princeton.edu/news/building-career-path-research-software-engineers}},
  note = {Accessed 2025-06-25} % Added access date
}

@misc{PawlickaDeger2022,
  author = {Pawlicka-Deger, Urszula},
  year = {2022},
  month = jul,
  title = {Digital humanities needs equality between humanists and technicians},
  journal = {Times Higher Education},
  howpublished = {\url{https://www.timeshighereducation.com/blog/digital-humanities-needs-equality-between-humanists-and-technicians}},
  note = {Accessed 2025-06-25} % Added access date
}

@incollection{Prentice2022,
  author = {Prentice, Sophie},
  year = {2022},
  title = {The Fear of Losing Your Job},
  booktitle = {The Future of Workplace Fear: How Human Reflex Stands in the Way of Digital Transformation},
  pages = {81-97},
  publisher = {Apress},
  OPTaddress ={Berkeley, CA}
}

@incollection{Reicher2010,
  author = {Reicher, Stephen D. and Spears, Russell and Haslam, S. Alexander},
  year = {2010},
  title = {The social identity approach in social psychology},
  booktitle = {The SAGE handbook of identities},
  OPTeditor ={Wetherell, Margaret and Mohanty, Chandra T.},
  pages = {45-62},
  publisher = {SAGE},
  OPTaddress ={London}
}

@article{Reynolds2001,
  author = {Reynolds, Katherine J. and Turner, John C. and Haslam, S. Alexander and Ryan, Michelle K.},
  year = {2001},
  title = {The role of personality and group factors in explaining prejudice},
  journal = {J. of Experimental Social Psychology},
  volume = {37},
  pages = {427-434}
}

@article{Riker1968,
  author = {Riker, William H. and Ordeshook, Peter C.},
  year = {1968},
  title = {A Theory of the Calculus of Voting},
  journal = {American political science review},
  volume = {62},
  number = {1},
  pages = {25-42}
}

@article{Rosso2010,
  author = {Rosso, Brian D. and Dekas, Kathryn H. and Wrzesniewski, Amy},
  year = {2010},
  title = {On the meaning of work: A theoretical integration and review},
  journal = {Research in organizational behavior},
  volume = {30},
  pages = {91-127}
}

@article{SeymourSmith2017,
  author = {Seymour-Smith, Megan and Cruwys, Tegan and Haslam, S. Alexander and Brodribb, Warwick},
  year = {2017},
  title = {Loss of group memberships predicts depression in postpartum mothers},
  journal = {Social Psychiatry and Psychiatric Epidemiology},
  volume = {52},
  pages = {201-210}
}

@article{Sharma2017,
  author = {Sharma, Pushkal Kumar and Misra, Rama Kant and Mishra, Praveen},
  year = {2017},
  title = {Job satisfaction scale: adaptation and validation among Indian IT (information technology) employees},
  journal = {Global Business Review},
  volume = {18},
  number = {3},
  pages = {703-718}
}

@article{Shih1999,
  author = {Shih, Margaret and Pittinsky, Todd L. and Ambady, Nalini},
  year = {1999},
  title = {Stereotype susceptibility: Identity salience and shifts in quantitative performance},
  journal = {Psychological Science},
  volume = {10},
  pages = {80-83}
}

@misc{Smith2014,
  author = {Smith, Alan},
  year = {2014},
  month = jul,
  title = {Research Fortnight. Engineering a future for research software and its makers},
  howpublished = {\url{https://www.researchresearch.com/news/article/?articleId=1345478}},
  note = {Accessed 2025-06-25} % Added access date
}

@article{Steffens2017,
  author = {Steffens, Niklas K. and Haslam, S. Alexander and Schuh, Sebastian C. and Jetten, Jolanda and van Dick, Rolf},
  year = {2017},
  title = {A meta-analytic review of social identification and health in organizational contexts},
  journal = {Personality and Social Psychology Review},
  volume = {21},
  number = {4},
  pages = {303-335}
}

@article{Steger2012,
  author = {Steger, Michael F. and Dik, Bryan J. and Duffy, Ryan D.},
  year = {2012},
  title = {Measuring meaningful work: The work and meaning inventory (WAMI)},
  journal = {J. of career Assessment},
  volume = {20},
  number = {3},
  pages = {322-337}
}

@article{Subasic2008,
  author = {Subašić, Eva and Reynolds, Katherine J. and Turner, John C.},
  year = {2008},
  title = {The political solidarity model of social change: Dynamics of self-categorization in intergroup power relations},
  journal = {Personality and Social Psychology Review},
  volume = {12},
  number = {4},
  pages = {330-352}
}

@incollection{Tajfel1979,
  author = {Tajfel, Henri and Turner, John C.},
  year = {1979},
  title = {An integrative theory of intergroup conflict},
  booktitle = {The social psychology of intergroup relations},
  OPTeditor ={Austin, W. G. and Worchel, S.},
  pages = {33-37},
  publisher = {Brooks/Cole},
  OPTaddress ={Monterey, CA}
}

@phdthesis{Toepfer1996,
  author = {Toepfer, Elizabeth A.},
  year = {1996},
  title = {The career satisfaction and success of corporate executives: The relationship among attachment styles, sex-type, and gender},
  school = {Teachers College, Columbia University}
}

@book{Turner1987,
  author = {Turner, John C. and Hogg, Michael A. and Oakes, Penelope J. and Reicher, Stephen D. and Wetherell, Margaret S.},
  year = {1987},
  title = {Rediscovering the social group: A self-categorization theory},
  publisher = {Basil Blackwell},
  OPTaddress ={Cambridge, MA}
}

@article{VanZomeren2012,
  author = {Van Zomeren, Martijn and Leach, Colin W. and Spears, Russell},
  year = {2012},
  title = {Protesters as “passionate economists” a dynamic dual pathway model of approach coping with collective disadvantage},
  journal = {Personality and Social Psychology Review},
  volume = {16},
  number = {2},
  pages = {180-199}
}

@article{VanZomeren2004,
  author = {Van Zomeren, Martijn and Spears, Russell and Fischer, Agneta H. and Leach, Colin W.},
  year = {2004},
  title = {Put your money where your mouth is! Explaining collective action tendencies through group-based anger and group efficacy},
  journal = {J. of personality and social psychology},
  volume = {87},
  number = {5},
  pages = {649}
}

@article{Wakslak2007,
  author = {Wakslak, Cheryl J. and Jost, John T. and Tyler, Tom R. and Chen, Ellen S.},
  year = {2007},
  title = {Moral outrage mediates the dampening effect of system justification on support for redistributive social policies},
  journal = {Psychological science},
  volume = {18},
  number = {3},
  pages = {267-274}
}

@incollection{Wegge2014,
  author = {Wegge, Jürgen and Halsam, S. Alexander},
  year = {2014},
  title = {Group goal setting, social identity, and self-categorization: Engaging the collective self to enhance group performance and organizational outcomes},
  booktitle = {Social identity at work: Developing theory for organizational practice (2nd ed.)},
  OPTeditor ={Haslam, S. Alexander and van Knippenberg, Daan and Platow, Michael and Ellemers, Naomi},
  pages = {43-60},
  publisher = {Psychology Press},
  OPTaddress ={Hove, UK}
}

@techreport{Boyd2022,
  author = {Boyd, R. L. and Ashokkumar, A. and Seraj, S. and Pennebaker, J. W.},
  title = {The development and psychometric properties of {LIWC}-22},
  institution = {University of Texas at Austin},
  OPTaddress ={Austin, TX},
  year = {2022},
  url = {https://www.liwc.app}
}

@incollection{Boyd2017,
  author = {Boyd, R. L.},
  title = {Psychological text analysis in the digital humanities},
  booktitle = {Data Analytics in Digital Humanities},
  OPTeditor ={Hai-Jew, S.},
  publisher = {Springer International Publishing},
  pages = {161--189},
  year = {2017},
  OPTdoi = {10.1007/978-3-319-54499-1_7}
}

@article{tingley2014mediation,
  title={Mediation: {R} package for causal mediation analysis},
  author={Tingley, Dustin and Yamamoto, Teppei and Hirose, Kentaro and Keele, Luke and Imai, Kosuke},
  journal={J. of statistical software},
  volume={59},
  pages={1--38},
  year={2014}
}

@article{segal2008developing,
  title={Developing scientific software},
  author={Segal, Judith and Morris, Chris},
  journal={IEEE software},
  volume={25},
  number={4},
  pages={18--20},
  year={2008},
  publisher={IEEE}
}

@inproceedings{carver2007software,
  title={Software development environments for scientific and engineering software: A series of case studies},
  author={Carver, Jeffrey C and Kendall, Richard P and Squires, Susan E and Post, Douglass E},
  booktitle={29th Int'l Conf. on Software Engineering},
  pages={550--559},
  year={2007},
  organization={IEEE}
}

@misc{marion2022modelling,
  title={Modelling: Understanding pandemics and how to control them. Epidemics 39, 100588},
  author={Marion, G and Hadley, L and Isham, V and Mollison, D and Panovska-Griffiths, J and Pellis, L and Tomba, GS and Scarabel, F and Swallow, B and Trapman, P and others},
  year={2022}
}

@article{chung2008revealing,
  title={Revealing dimensions of thinking in open-ended self-descriptions: An automated meaning extraction method for natural language},
  author={Chung, Cindy K and Pennebaker, James W},
  journal={J. of research in personality},
  volume={42},
  number={1},
  pages={96--132},
  year={2008},
  publisher={Elsevier}
}

@article{back2018we,
  title={From I to we: Group formation and linguistic adaption in an online xenophobic forum},
  author={B{\"a}ck, Emma A and B{\"a}ck, Hanna and Send{\'e}n, Marie Gustafsson and Sikstr{\"o}m, Sverker},
  journal={J. of Social and Political Psychology},
  volume={6},
  number={1},
  pages={76--91},
  year={2018}
}

@inproceedings{Franca2014, 
author = {Fran\c{c}a, C\'{e}sar and Sharp, Helen and da Silva, Fabio Q. B.}, 
title = {Motivated software engineers are engaged and focused, while satisfied ones are happy}, 
year = {2014}, 
booktitle = {8th Int'l Symposium on Empirical Software Engineering and Measurement}, 
articleno = {32}, 
numpages = {8}, 
OPTlocation = {Torino, Italy}, 
series = {ESEM '14},
publisher = {ACM},
}

@article{Graziotin2019,
  author       = {Daniel Graziotin and
                  Fabian Fagerholm},
  title        = {Happiness and the productivity of software engineers},
  journal      = {CoRR},
  volume       = {abs/1904.08239},
  year         = {2019}
}

@ARTICLE{França2020,
  author={França, César and da Silva, Fabio Q. B. and Sharp, Helen},
  journal={IEEE Trans. on Software Engineering}, 
  title={Motivation and Satisfaction of Software Engineers}, 
  year={2020},
  volume={46},
  number={2},
  pages={118-140}
}

@ARTICLE{Hovenden1996,
author={Hovenden, F.M. and Walker, S.D. and Sharp, H.C. and Woodman, M.},
title={Building quality into scientific software}, 
journal={Software Quality}, 
year={1996},
volume={5},
pages={25-32}
}

@inproceedings{Storey2022, 
author = {Storey, Margaret-Anne and Houck, Brian and Zimmermann, Thomas}, 
title = {How developers and managers define and trade productivity for quality}, 
year = {2022}, 
booktitle = {15th Int'l Conf. on Cooperative and Human Aspects of Software Engineering}, 
pages = {26–35}, 
numpages = {10}, 
OPTlocation = {Pittsburgh, Pennsylvania}, 
OPTseries = {CHASE '22} 
}

@INPROCEEDINGS{Rauf2025,
  author={Rauf, Irum and Sharp, Helen and Lopez, Tamara and Wermelinger, Michel},
  booktitle={18th Int'l Conf. on Cooperative and Human Aspects of Software Engineering}, 
  title={Human-Machine Teaming and Team Effectiveness in AI Tools for Software Engineering}, 
  year={2025},
  publisher={IEEE},
  number={},
  pages={75-80}
}

@article{Breaugh1985,
author = {James A. Breaugh},
title ={The Measurement of Work Autonomy},
journal = {Human Relations},
volume = {38},
number = {6},
pages = {551-570},
year = {1985}
}

@inproceedings{ICSE2008workshop, 
author = {Carver, Jeffrey C.}, 
title = {{SE-CSE} 2008: the first Int'l workshop on software engineering for computational science and engineering}, 
year = {2008}, 
publisher = {ACM}, 
booktitle = {Companion of the 30th Int'l Conf. on Software Engineering}, 
pages = {1071–1072}, 
OPTlocation = {Leipzig, Germany}, 
OPTseries = {ICSE Companion '08} 
}

@Article{allenetal2017,
  author =	{Allen, Alice and Aragon, Cecilia and Becker, Christoph and Carver, Jeffrey and Chis, Andrei and Combemale, Benoit and Croucher, Mike and Crowston, Kevin and Garijo, Daniel and Gehani, Ashish and Goble, Carole and Haines, Robert and Hirschfeld, Robert and Howison, James and Huff, Kathryn and Jay, Caroline and Katz, Daniel S. and Kirchner, Claude and Kuksenok, Katie and L\"{a}mmel, Ralf and Nierstrasz, Oscar and Turk, Matt and van Nieuwpoort, Rob and Vaughn, Matthew and Vinju, Jurgen J.},
  title =	{{Engineering Academic Software (Dagstuhl Perspectives Workshop 16252)}},
  pages =	{1--20},
  journal =	{Dagstuhl Manifestos},
  ISSN =	{2193-2433},
  year =	{2017},
  volume =	{6},
  number =	{1},
  publisher =	{Schloss Dagstuhl -- Leibniz-Zentrum f{\"u}r Informatik},
  address =	{Dagstuhl, Germany},
}

@ARTICLE{Dingsøyr2022,
  author={Dingsøyr, Torgeir and Jørgensen, Magne and Carlsen, Frode Odde and Carlström, Lena and Engelsrud, Jens and Hansvold, Kine and Heibø-Bagheri, Mari and Røe, Kjetil and Sørensen, Karl Ove Vika},
  journal={IT Professional}, 
  title={Enabling Autonomous Teams and Continuous Deployment at Scale}, 
  year={2022},
  volume={24},
  number={6},
  pages={47-53}
 }

@misc{lassenius2025,
      title={Towards a Taxonomy for Autonomy in Large-Scale Agile Software Development}, 
      author={Casper Lassenius and Torgeir Dingsøyr},
      year={2025},
      eprint={2503.02651},
      archivePrefix={arXiv},
      primaryClass={cs.SE},
      url={https://arxiv.org/abs/2503.02651}, 
}

@ARTICLE{Hoda2013,
  author={Hoda, Rashina and Noble, James and Marshall, Stuart},
  journal={IEEE Trans. on Software Engineering}, 
  title={Self-Organizing Roles on Agile Software Development Teams}, 
  year={2013},
  volume={39},
  number={3},
  pages={422-444}
  }

@INPROCEEDINGS{acharya2019,
  author={Acharya, Bibek and Colomo-Palacios, Ricardo},
  booktitle={19th Int'l Conf. on Computational Science and Its Applications}, 
  title={A Systematic Literature Review on Autonomous Agile Teams}, 
  year={2019},
  pages={146-151}
  }

@book{Ryan2017,
  author = {Ryan, RM and Deci, EL},
  year = {1979},
  title = {Self-determination theory. Basic psychological needs in motivation, development, and wellness},
    publisher = {The Guildford Press}
}

@ARTICLE{procaccino2005,
  author={Procaccino, J. Drew and Verner, June M. and Shelfer, Katherine M. and Gefen, David},
  journal={J. of Systems and Software}, 
  title={What do software practitioners really think about project success: an exploratory study}, 
  year={2005},
  volume={78},
  number={2},
  pages={194-203}
  }

@misc{RSEcredit,
  author = {Druskat, Stephan and Katz, Daniel S. and  Klein, David and  Santcroos, Mark and  Schlauch, Tobias and  Sexton-Kennedy, Liz and  Truskinger, Anthony},
  title = {Policy},
  howpublished = {\url{https://www.software.ac.uk/blog/credit-and-recognition-research-software-current-state-practice-and-outlook}},
  note = {Accessed 2025-10-23} 
}

@inproceedings{JCT1991, 
author = {Sein, Maung K. and Bostrom, Robert P.}, 
title = {A psychometric study of the job characteristics scale of the job diagnostic survey in an MIS setting}, 
year = {1991}, 
publisher = {ACM}, 
booktitle = {Conf. on SIGCPR}, 
pages = {96–110}, 
OPTseries = {SIGCPR '91},
}

@inproceedings{trinkenreichvirtual2022, 
author = {Trinkenreich, Bianca and Stol, Klaas-Jan and Sarma, Anita and German, Daniel M. and Gerosa, Marco A. and Steinmacher, Igor}, 
title = {Do I Belong? Modeling Sense of Virtual Community Among Linux Kernel Contributors}, 
year = {2023}, 
publisher = {IEEE Press}, 
booktitle = {Proceedings of the 45th ICSE}, 
pages = {319–331}
}

\end{document}